\documentclass[pra,twocolumn,showpacs,superscriptaddress,floatfix, reprint]{revtex4-2}
\usepackage[usenames,dvipsnames]{color}
\usepackage[latin1]{inputenc}
\usepackage[english]{babel}
\usepackage{graphicx}
\usepackage{color,soul}
\usepackage{amssymb,amsmath}
\usepackage[Gray,squaren]{SIunits}
\usepackage{xspace}
\usepackage{upgreek}
\usepackage{ulem}
\usepackage{epstopdf}
\usepackage{amssymb,amsmath,verbatim,ulem}
\usepackage{hyperref}

\usepackage{chngcntr}
\usepackage[toc,page]{appendix}

\newcommand{\ie}{i.e.\@\xspace}

\newcommand{\fig}[1]{Fig.~\ref{fig:#1}}

\newcommand{\ket}[1]{|{#1}\rangle}

\begin{document}

\title{Homodyne detection of a two-photon resonance assisted by cooperative emission}
\author{Chetan Sriram Madasu}
\affiliation{Quantum Hub, Division of Physics and Applied Physics, Nanyang Technological University, 21 Nanyang Link, 637371 Singapore}
\author{Chang Chi Kwong}
\affiliation{Quantum Hub, Division of Physics and Applied Physics, Nanyang Technological University, 21 Nanyang Link, 637371 Singapore}
\affiliation{MajuLab, CNRS-UCA-SU-NUS-NTU International Joint Research Unit, Singapore}
\author{David Wilkowski}
\email{david.wilkowski@ntu.edu.sg}
\affiliation{Quantum Hub, Division of Physics and Applied Physics, Nanyang Technological University, 21 Nanyang Link, 637371 Singapore}
\affiliation{MajuLab, CNRS-UCA-SU-NUS-NTU International Joint Research Unit, Singapore}
\affiliation{Centre for Quantum Technologies, National University of Singapore, 3 Science Drive 2, 117543 Singapore}
\affiliation{Centre for Disruptive Photonic Technologies, The Photonics Institute, Nanyang Technological University, Singapore 637371, Singapore}
\author{Kanhaiya Pandey}
\email{kanhaiyapandey@iitg.ac.in}
\affiliation{Department of Physics, Indian Institute of Technology Guwahati, Guwahati, Assam 781039, India }

\date{\today{}}

\begin{abstract}

Focusing on the transient regime, we explore atomic two-photon spectroscopy with self-aligned homodyne interferometry in a $\Lambda$-system with large optical depth. The two light sources at the origin of the interference are the single-photon transient transmission of the probe, and the slow light of the electromagnetically induced transparency. By switching off the probe laser abruptly (flash effect), the transient transmission signal is reinforced by cooperativity, showing enhanced sensitivity to the two-photon frequency detuning. If the probe laser is periodically switched on and off, the amplitude of the transmission signal varies and remains large even for high modulation frequency. This technique has potential applications in sensing, such as magnetometry and velocimetry, and in coherent population trapping clocks.

\end{abstract}


\maketitle

\section{Introduction}
Since the first experimental evidence of coherent population trapping (CPT) \cite{alzetta1976experimental} and electromagnetically induced transparency (EIT) \cite{boller1991observation}, the narrow dark two-photon resonance \cite{arimondo1976nonabsorbing} has been a key feature for many fundamental and practical phenomena. Without being exhaustive, we point out: slowing and storage of light \cite{HHD99, KAF18}, quantum memories \cite{BSA13,HTC18}, CPT clocks \cite{MXT16, V05, LIY17, LMJ13,LYT17, RCX20}, microwave and terahertz generation and detection \cite{JSA12,SNP18,WML18,lam2021directional}, laser cooling \cite{aspect1988laser,Wilkowski_2009}, Raman velocimetry \cite{CLH20}, and many-body photonics systems \cite{FIM05,tebben2021nonlinear,firstenberg2013attractive,carusotto2013quantum}. A generalization of CPT, involving dark multi-photon resonances, has also been demonstrated recently \cite{PhysRevApplied.12.034035}. 

Various methods have been proposed and implemented to narrow down the two-photon resonance. A straightforward approach consists in increasing the single-photon optical density (OD) of the medium. However, the linewidth of the two-photon resonance decreases as the square root of OD \cite{LFZ97}. At large OD, interference between the probe field and a generated field due to enhanced coherent Raman scattering further leads to a linewidth narrowing~\cite{LFZ97, GLM02}. Here, the linewidth reduction is more pronounced since it is inversely proportional to the OD. However, these two fields have a large frequency difference (in the range of  ground-state hyperfine splitting). Hence, this technique relies on a heterodyne detection method leading to an asymmetry of the lineshape, which causes systematic frequency shifts \cite{LFZ97}. Finally, narrowing of the two-photon resonance can also be achieved using cooperative effects, for example, by embedding the atomic medium into an optical resonator \cite{BLL00}. Using this method, the linewidths of \textit{bad} cavities have been significantly narrowed down \cite{WGB00, LFS98,HZZ07,WGX08}. One notes that the narrowing of the EIT dip using atoms inside a cavity induces frequency pulling, which also causes systematic shifts of the EIT resonance \cite{LFS98,HZZ07}.

The state-of-the-art CPT clock is based upon the time-separated Ramsey pulse method \cite{THE82, PST08, ZGC05, barantsev2018line,EBI13,BRD15,LYT17}. Here, a large OD helps to improve the overall signal-to-noise ratio. The time-separated Ramsey pulse method has been implemented in vapor cells \cite{THE82, PST08, ZGC05, barantsev2018line} and cold atomic ensembles \cite{EBI13,BRD15,LYT17}. Ultimately, the linewidth of the two-photon resonance scales like $1/T$, where $T$ is the free evolution time between the two (short) Ramsey pulses. Otherwise, the linewidth is limited by the decoherence rate between the two ground-state hyperfine levels. This decoherence rate is generally larger for vapor cells than for cold atoms systems.

In this paper, we propose a new approach to measure a two-photon resonance using transient and cooperative effects in a dilute EIT medium with large OD. By switching on and off the probe beam once or several times, we extract a phase-sensitive signal between the single-photon and the EIT transmitted fields (see Fig. \ref{schematics}a). Our proposal does not entail the time-separated Ramsey pulse technique. Instead of measuring the interference between the optical field and the free evolution of the atomic coherence, we measure the interference between two self-aligned optical fields; namely the single-photon and the EIT transmitted fields. The linewidth of the two-photon resonance is inversely proportional to the OD of the medium, as in Ref. \cite{LFZ97}.

Our approach leads to two major improvements over previous techniques. First, cooperativity can enhance the transient signal up to four times the incident field, as observed in \textit{flash} experiments \cite{KYP14}. Second, the different time scales in the atomic system allow for high-repetition-rate homodyne interference between the single-photon and the EIT transmitted fields. 

\section{Model and Results}
\begin{figure}
    \centering
    \includegraphics[scale=0.6]{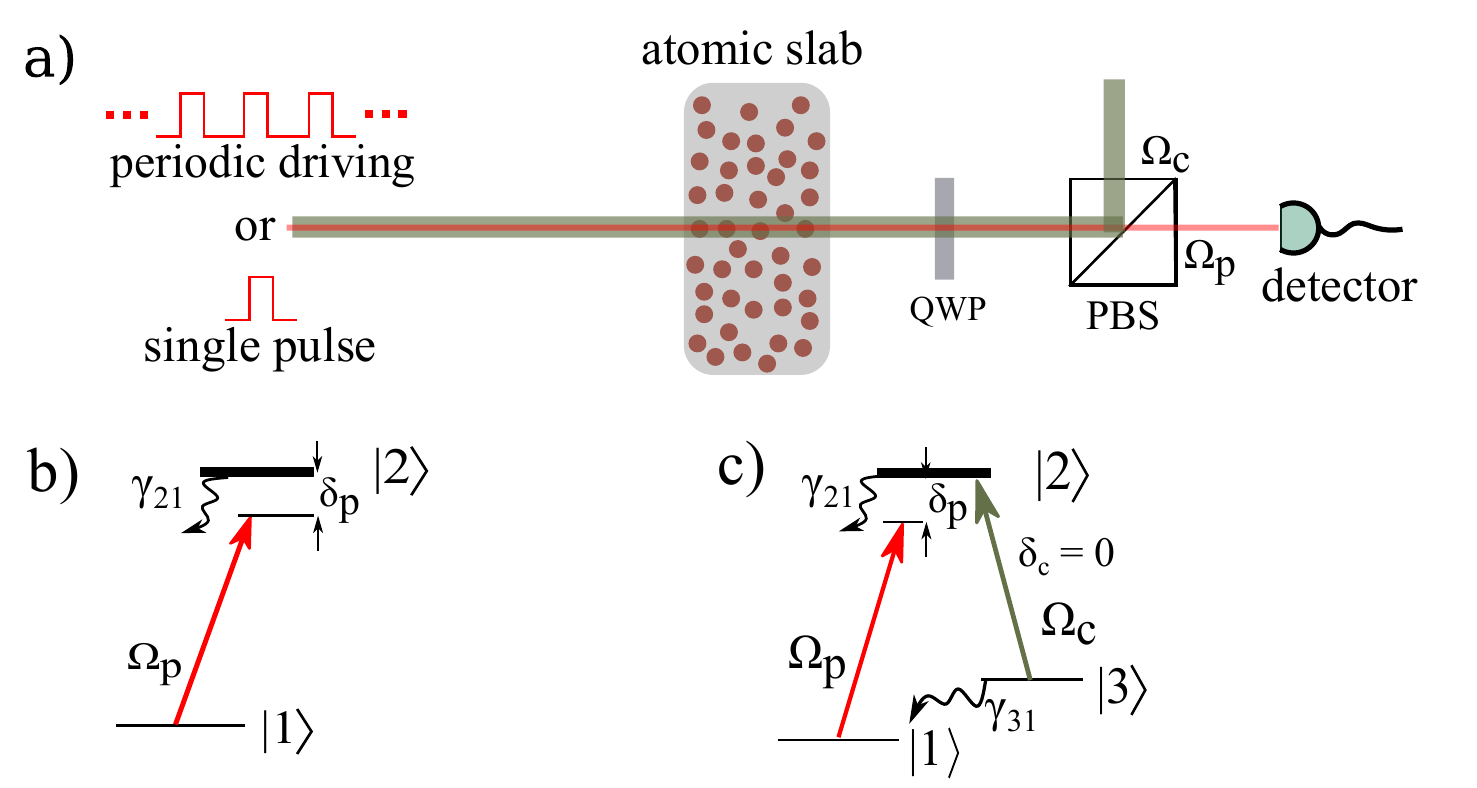}
    \caption{(a) Two plane waves, a control beam and a probe beam, are resonantly coupled to a slab of atomic $\Lambda$-system. The probe field is abruptly switched off or periodically turned on and off. Its transmitted intensity is measured on a detector. (b $\&$ c) Energy level schematics showing the active optical fields of the two-level and three-level systems, respectively. Here, $\Omega_c$($\Omega_p$) and $\delta_c$($\delta_p$) are the control(probe) field Rabi frequency and detuning, respectively. $\gamma_{21}$ is the excited state linewidth and $\gamma_{31}/2$ is the ground state decoherence rate.}
    \label{schematics}
\end{figure}

We now describe the model and the assumptions used in our study.  We consider a plane wave probe beam shining on a slab of zero temperature atomic medium with a thickness $L$. For a weak probe beam below saturation, the transmitted field  at time $t$ is $E(t)\exp(-i\omega_pt)$, where the complex amplitude reads~\cite{WCL09,KYP14}
\begin{equation}
E(t)=\frac{1}{2\pi}\int^{\infty}_{-\infty} \tilde{E}_i (\omega) \exp\left[-\frac{b(\delta)}{2}+i\phi(\delta)\right]e^{-i(\omega-\omega_p) t}d\omega.
\label{fft}
\end{equation}
 $\tilde{E}_i(\omega)$ is the Fourier component of the incident field at a frequency $\omega$,  $\delta = \omega-(E_2-E_1)/\hbar$ is the frequency detuning of the Fourier component relative to the resonance of the $\ket{1}\rightarrow\ket{2}$ probe transition  (see Fig. \ref{schematics}b), and $\omega_p$ is the carrier frequency of the probe beam. $b(\delta) = \textrm{Im}\{\chi(\delta)\} kL$ and
$\phi(\delta) = \textrm{Re}\{\chi(\delta)\}kL/2$ are the OD and the phase shift for an optical field at detuning $\delta$, respectively. \\

\subsection{Two-level system}
As a preliminary study, we consider a two-level system where
\begin{equation}
    \chi(\delta)\equiv\chi_{1p}(\delta)=\frac{-6\pi\rho}{k^3}\frac{\gamma_{21}}{2}\left(\delta+\frac{i\gamma_{21}}{2}\right)^{-1}
\end{equation}
is the one-photon susceptibility of the dilute atomic ensemble assuming each atom scatters light independently. Here, $\gamma_{21}$ is the transition linewidth, $\rho$ is the spatial density of the medium, and $k$ is the wave-number of the optical field. The bulk properties of the medium, namely $\rho$ and $L$, are encapsulated into a single parameter, $b_0\equiv b(0) = \frac{6\pi\rho L}{k^2}$, which is the resonant single-photon OD. We assume that $b_0\gg1$ and disregard the incoherent multiple scattering field in the forward transmission~\cite{datsyuk2006diffuse}. \\

\begin{figure}
\includegraphics[scale=0.55]{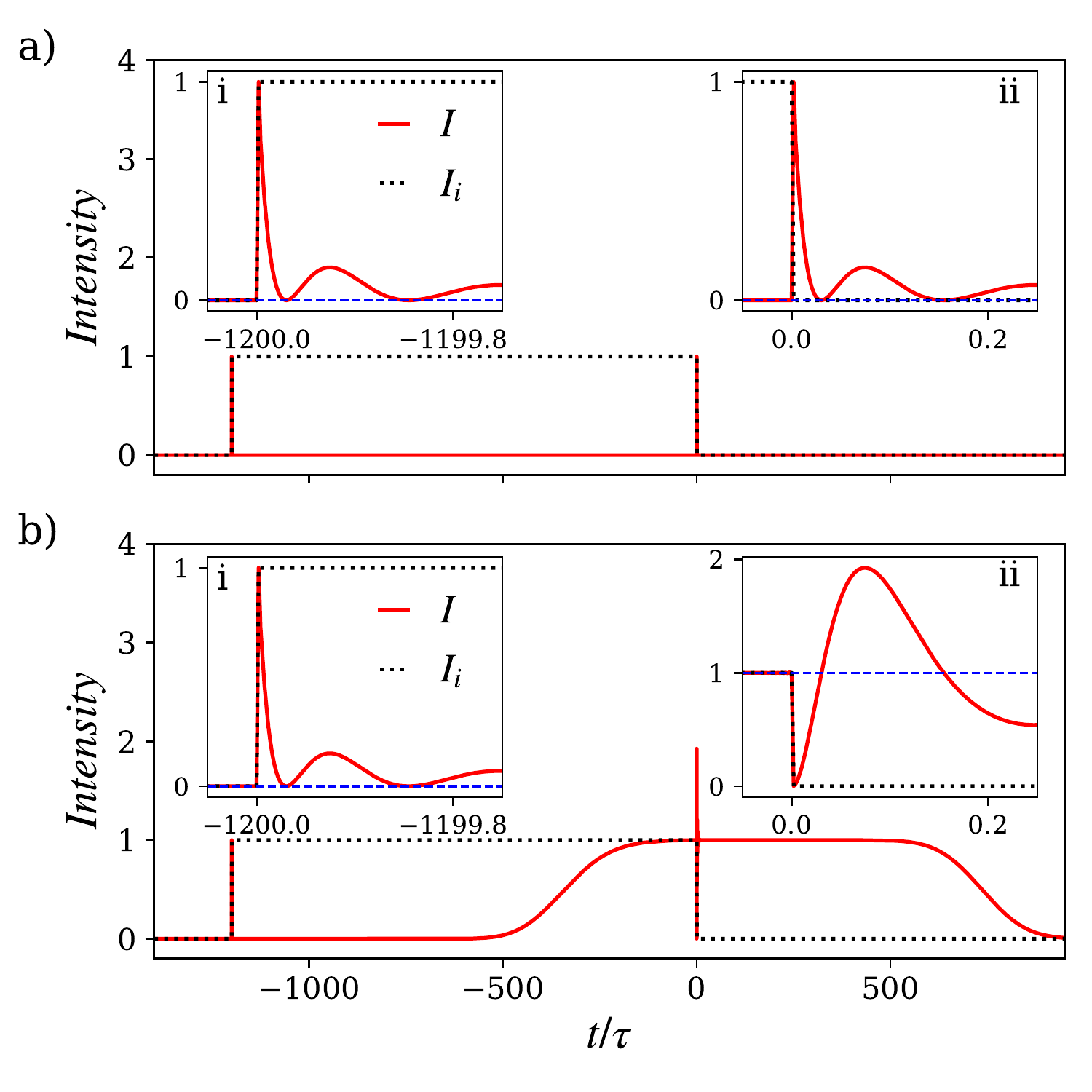}
\caption{\label{Intensity_vs_time} (a $\&$ b) Temporal evolution of the transmitted intensity (red curve) with a square input pulse (black dotted curve), for $b_0 = 200$. The SBO precursors at probe ignition and flashes at probe extinction are shown in the insets i and ii, respectively. The dashed blue lines are the intensity levels that the flashes or precursors relax to. (a) Two-level system   i.e., $\Omega_c=0$. (b) Three-level system with $\Omega_{c} = \gamma_{21}/2$, $\gamma_{31}=0$ and $\delta_c=0$. }
\end{figure}

At first, the probe beam is turned on abruptly and shined for a duration long enough for the steady-state regime to be established, before being turned off abruptly  (see the black dotted curve in Fig. \ref{Intensity_vs_time}a). Defining $\delta_p$ as the probe beam carrier detuning, the resonant ($\delta_p=0$) transmitted intensity, $I(t)\equiv|E(t)|^2$, is illustrated by the red curve in Fig. \ref{Intensity_vs_time}a. Two flashes of light are observed. The initial flash (see inset i  in Fig. \ref{Intensity_vs_time}a), during probe ignition, corresponds to the Sommerfeld-Brillouin optical (SBO) precursors \cite{SOM14,BRI14,WCL09,BFY14,PYC12}. During probe extinction, the flash of light is related to the free induction decay (FID) signal \cite{BPP48,TTI97} (see inset ii in Fig. \ref{Intensity_vs_time}a). The flashes have fast cooperative decay times that scale like \cite{KYD15}
\begin{equation}
	\label{tau_f}
	\tau_f\sim\frac{1}{b_0\gamma_{21}}.
\end{equation}
For an optically thick medium, $\tau_f$ is much shorter than excited state lifetime $\tau=\gamma_{21}^{-1}$. \\

To gain physical insight into the flashes, we decompose the amplitude of the transmitted field as
\begin{equation}
	\label{E_twolevel}
	E(t)=E_i(t)+E_s(t),
\end{equation}
where $E_i(t)$ and $E_s(t)$ are the amplitudes of the incident and forward scattered fields, respectively, in complex notation. In the steady-state regime, $E\approx 0$ means that $E_s\approx -E_i\approx -1$, \ie  a destructive interference between the incident field and the forward scattered field. Here, we use the normalization: $|E_i|=1$ when the field is on, and the following notation convention: when quantities do not explicitly depend on time, they correspond to the steady-state values. When the incident field is switched off at $t=0^+$, since the atoms have a finite response time, $E(0^+)\approx E_s\approx -1$ and $I(0^+)\approx 1$ (see  Fig. \ref{Intensity_vs_time}a). As pointed out in Refs.~\cite{CPD11,KYP14}, the FID flash is a measurement of $E_s$.\\

The FID flash and SBO precursors in Fig. \ref{Intensity_vs_time}a have the same temporal profile. This can be understood from the temporal version of the Babinet principle~\cite{J99}. For $\delta_p=0$ and $b_0\gg1$, the transmitted fields during probe ignition and extinction should sum to zero, \ie, the steady-state transmitted field value~\cite{CPD11}. In other words, the FID flash and SBO precursor fields are complementary, \ie, they have equal amplitudes but opposite signs at resonance.\\

\subsection{Three-level $\Lambda$ system}

We now turn to the case of the $\Lambda$-system, with an additional state $\ket{3}$. In the presence of  a continuous control beam on the transition $\ket{3} \rightarrow \ket{2}$ ($\Omega_c\neq0$, see Fig. \ref{schematics}b), the EIT phenomenon takes place and the susceptibility of the probe beam reads,
\begin{equation}
\label{3levelT}
\chi(\delta)=\chi_{1p}(\delta)\left[1 - \frac{|\Omega_{c}|^2/4}{\left(\delta + \frac{i\gamma_{21}}{2}\right)\left(\delta - \delta_{c}+\frac{i\gamma_{31}}{2}\right)}\right]^{-1},
\end{equation}
where $\gamma_{31}/2$ is the relaxation rate of coherence between the ground states $\ket{3}$ and $\ket{1}$, and $\delta_{c}$ is the frequency detuning of the control beam. Since the OD is large, the transmission signal weakly depends on $\delta_c$, and we set $\delta_{c}=0$. Therefore, $\delta_p=0$ also indicates the two-photon resonance condition. Under this condition, the temporal evolution of the transmitted intensity is shown in Fig. 
\ref{Intensity_vs_time}b and can be directly compared to the two-level case of Fig. \ref{Intensity_vs_time}a. A salient difference is the increase of the transmission well after the SBO precursor ($t\sim-500\gamma_{21}^{-1}$), due to the development of the EIT slow light. For $\gamma_{31}\ll\Omega_c,\gamma_{21}$, this occurs after a group delay of (see the appendix)
\begin{equation}
\label{phase_group}
\tau_{EIT}=b_0\frac{\gamma_{21}}{|\Omega_c|^{2}}.
\end{equation}

As previously observed with a rubidium cold gas \cite{WCL09}, we find SBO precursors at probe ignition and an anti-flash at probe extinction. The timescale of the SBO precursors and the anti-flash in the EIT system is the same as the two-level case and is given by $\tau_f$. This fast timescale suggests that the SBO precursors and (anti-)flashes are effects solely related to the two-level response. At large optical thickness, we have $\tau_f\ll\tau\ll\tau_{EIT}$, with the medium being strongly absorbing at timescales between $\tau$ and $\tau_{EIT}$. Therefore, we have a clear timescale separation between EIT effects and those that are two-level in nature. The following decomposition of the transmitted field captures this timescale separation.
\begin{equation}
	\label{E_threelevel}
	E(t)=E_i(t)+E_s(t)+E_{EIT}(t).
\end{equation}
Here, $E_s(t)$ is the forward scattered field of the two-level system ($\Omega_c=0$). It consists of the high frequency or fast varying component of the transmitted light associated with the flash effect. $E_{EIT}(t)$ describes the slow dynamics associated with the central narrow EIT transparency window. There is a physical interest in this new field decomposition, because of the clear time scale separation between the flash and the slow EIT light, \ie, $\tau_f\ll\tau_{EIT}$. 

\begin{figure}
    \centering
    \includegraphics[scale=0.6]{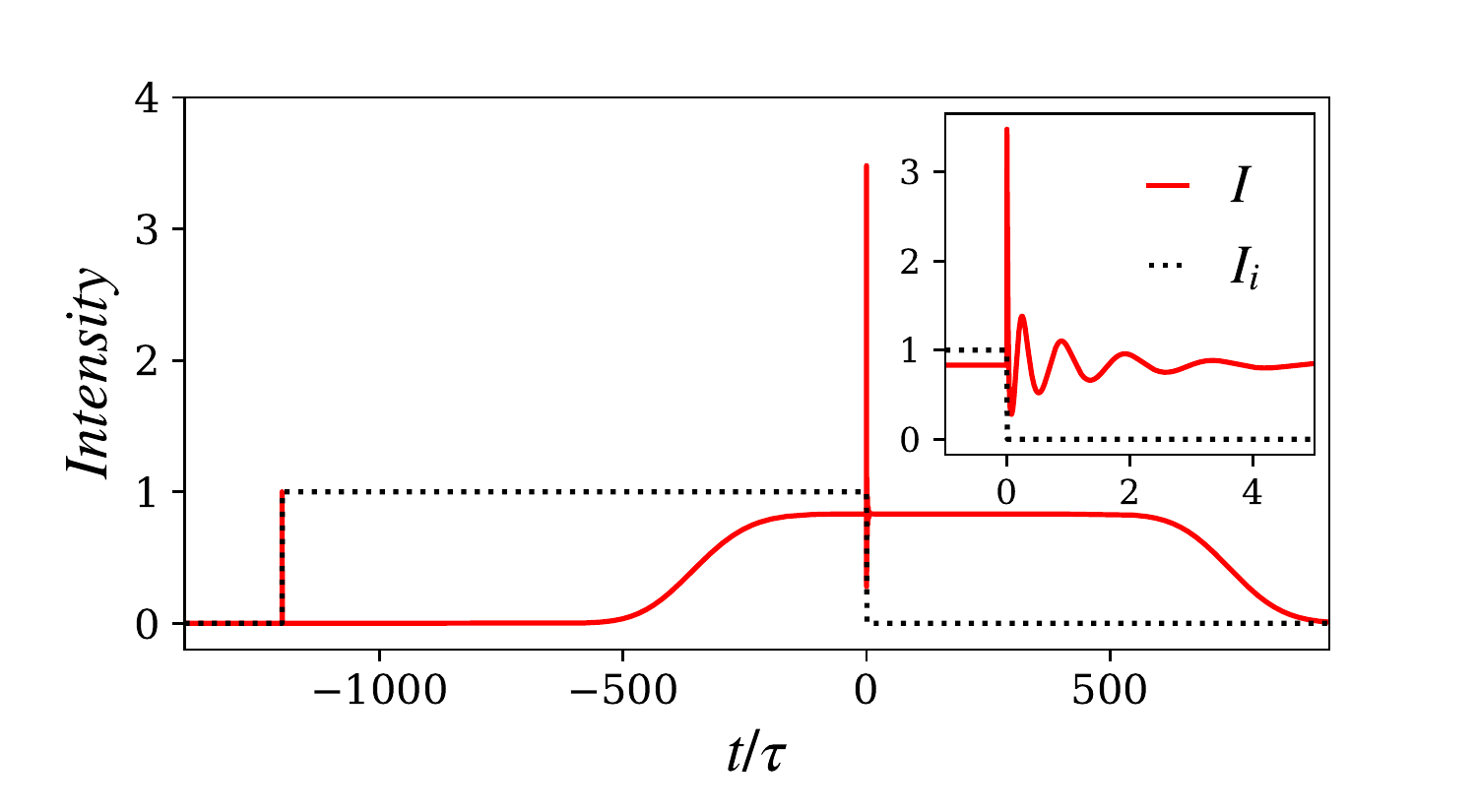}
    \caption{Transmission profile of the probe pulse (red curve) when the first maximum of $I(0^+)$ is reached for  $b_0=200$, $\Omega_c=\gamma_{21}/2, \delta_c=0$ and $\delta_p=\pi|\Omega_c|^2/({b_0}\gamma_{21})\approx3.8 \times 10^{-3}\gamma_{21}$. The black dotted curve is the incident pulse. The inset shows a zoomed-in view of  a superflash with $I(0^+)/I_i=3.5>1$ emitted during probe extinction.}
    \label{superflashInEIT}
\end{figure}

Since the SBO precursors originate from the two-level response of the probe transition, they are identical for both the two-level and three-level schemes, as shown in the insets i of Fig. \ref{Intensity_vs_time}a and Fig. \ref{Intensity_vs_time}b, respectively. By the complementary argument discussed before, the FID flashes as given by $E_s$, should also be identical for the two-level and three-level cases. Hence, the clear difference in the temporal profiles of the FID flashes (compare insets ii in Fig. \ref{Intensity_vs_time}a and Fig. \ref{Intensity_vs_time}b), solely originates from the interference with the additional $E_{EIT}$ term. This interference is well-captured by the field decomposition in Eq.~(\ref{E_threelevel}). At $\delta_p=0$ and $\gamma_{31}=0$, the three-level medium is transparent in the steady-state regime, meaning $E\approx 1$. Since $E_s\approx -1$ from the two-level case, we have $E_{EIT}\approx E\approx 1$. Thus $E(0^+)\approx E_s+E_{EIT}\approx 0$ during probe extinction, as observed in the inset ii of Fig. \ref{Intensity_vs_time}b. This anti-flash results from a destructive interference between the FID flash ($E_s$) and the EIT field ($E_{EIT}$).\\

\begin{figure}
\includegraphics[scale=0.6]{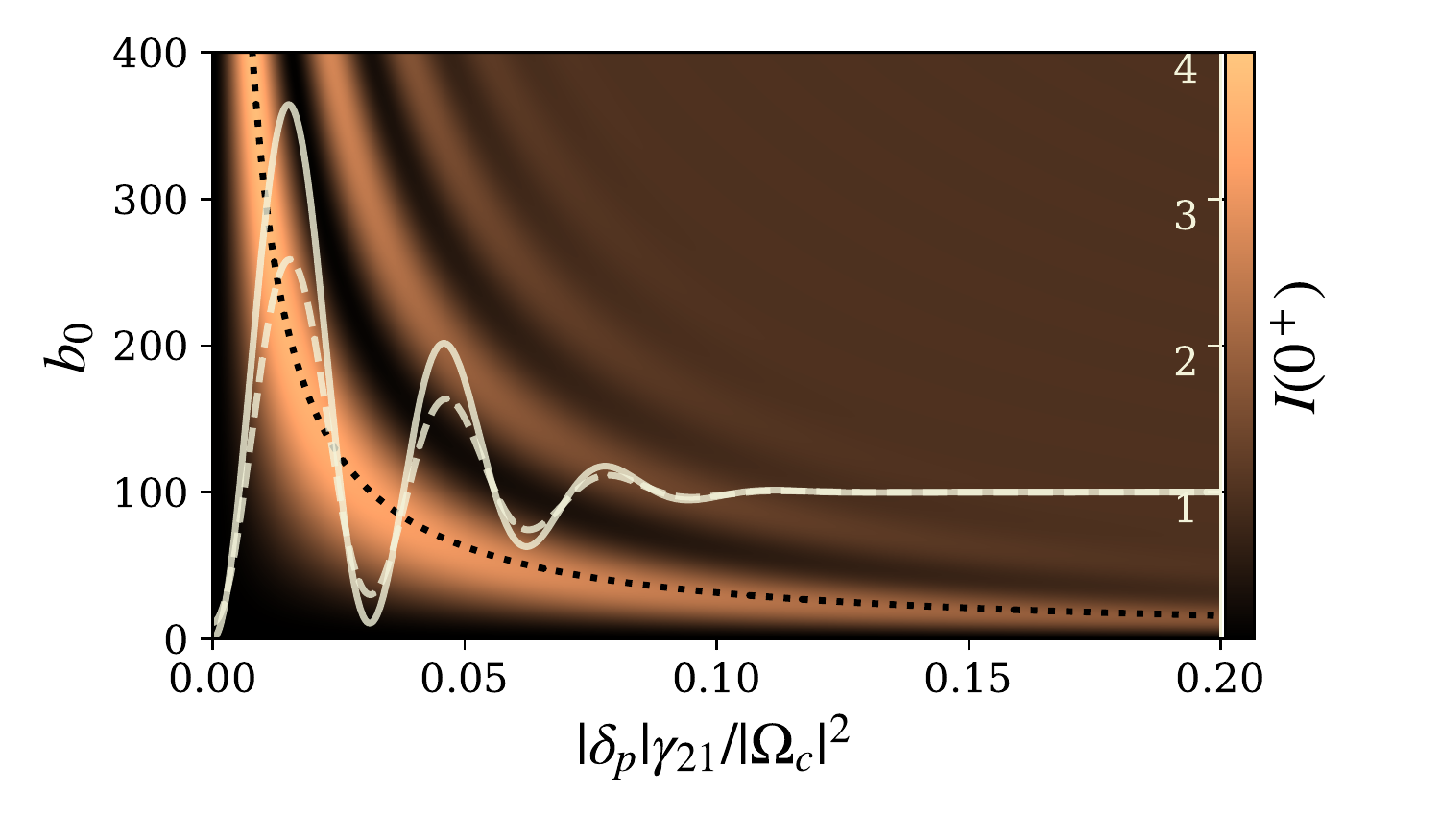}
\caption{\label{IIILevelIMaxWithOD} Color plot of the transmitted intensity at the probe extinction, $I(0^+)$, as functions of $b_0$ (left vertical axis) and the normalized probe detuning for $\Omega_{c} = \gamma_{21}/2$, $\delta_c=0$ and $\gamma_{31}=0$. The dotted black curve represents the first fringe maximum at a probe detuning of $\delta_{\pi}$, as defined in Eq. (\ref{detmax}). The light grey plain (dashed) curve, referred to the right vertical axis, is a section of the color plot for $b_0=200$, and $\gamma_{31}=0$ ($\gamma_{31}=\gamma_{21}/1000$).}
\end{figure}

The nature of the interference, constructive or destructive during probe extinction, depends upon the steady-state phase difference between $E_{EIT}$ and $E_s$:
\begin{equation}
	\label{Dphi}
	\Delta\phi=\phi_{EIT}-\phi_s  \approx \delta_p\tau_{EIT} - \pi = b_0\frac{\gamma_{21}}{|\Omega_c|^2}\delta_p-\pi.
\end{equation}
For a small probe detuning of $\delta_p\ll\gamma_{21}$, the phase of $E_s$ has an approximately constant value of $\phi_s\approx\pi$.
Neglecting off-resonance absorption inside the transparency window, the fringe maxima (constructive interference) is located at $\Delta\phi = 0 \,(\text{mod}\, 2\pi)$, \ie, $\phi_{EIT} = \pi\,(\text{mod}\, 2\pi)$.  This condition gives $E_{EIT}=E_s = -1$, and a maximum value of $I(0^+) = 4$, according to  Eq.~(\ref{E_threelevel}). In Fig. \ref{superflashInEIT}, we plot the transmission profile of the probe pulse in this condition, showing the constructive interference and the corresponding flash. The maximum of the flash does not reach the value of four as predicted, due to the presence of residual absorption on the $E_{EIT}$ field when the two-photon resonance condition is not exactly fulfilled. The transmission profile in Fig. \ref{superflashInEIT} is calculated at the first maxima where
\begin{equation}
	\label{detmax}
	|\delta_p|=\delta_\pi\equiv\frac{\pi}{b_0}\frac{|\Omega_c|^2}{\gamma_{21}}.
\end{equation}
The $b_0^{-1}$ prefactor in Eq. (\ref{detmax}) encapsulates the cooperative narrowing of the fringe pattern. We confirm this analysis by integrating Eq. (\ref{fft}) at $t=0^+$ with $\gamma_{31}=0$ for various values of $\delta_p$ and $b_0$ (see the 2D color plot in Fig. \ref{IIILevelIMaxWithOD}). The detuning at the first bright fringe is well captured by $\delta_\pi$ define in Eq.~(\ref{detmax}) (black dotted curve), and we clearly observed the enhancement of the phase measurement sensitivity, due to a cooperative narrowing of the fringes when $b_0$ increases.  \\

A section at $b_0=200$ is represented by the light grey curve in Fig. \ref{IIILevelIMaxWithOD}. As explained earlier, the first maximum does not reach the value of four due to the residual absorption at non-zero two-photon detunings. The absorption becomes more important at a larger detuning, which explains the damping of the fringes. The damping is reinforced when we include a non-zero relaxation of the ground state coherence, as illustrated for $\gamma_{31}=\gamma_{21}/1000$ by the light grey dashed curve in Fig. \ref{IIILevelIMaxWithOD}. 

\begin{figure}
    \flushleft
    \includegraphics[scale=0.55]{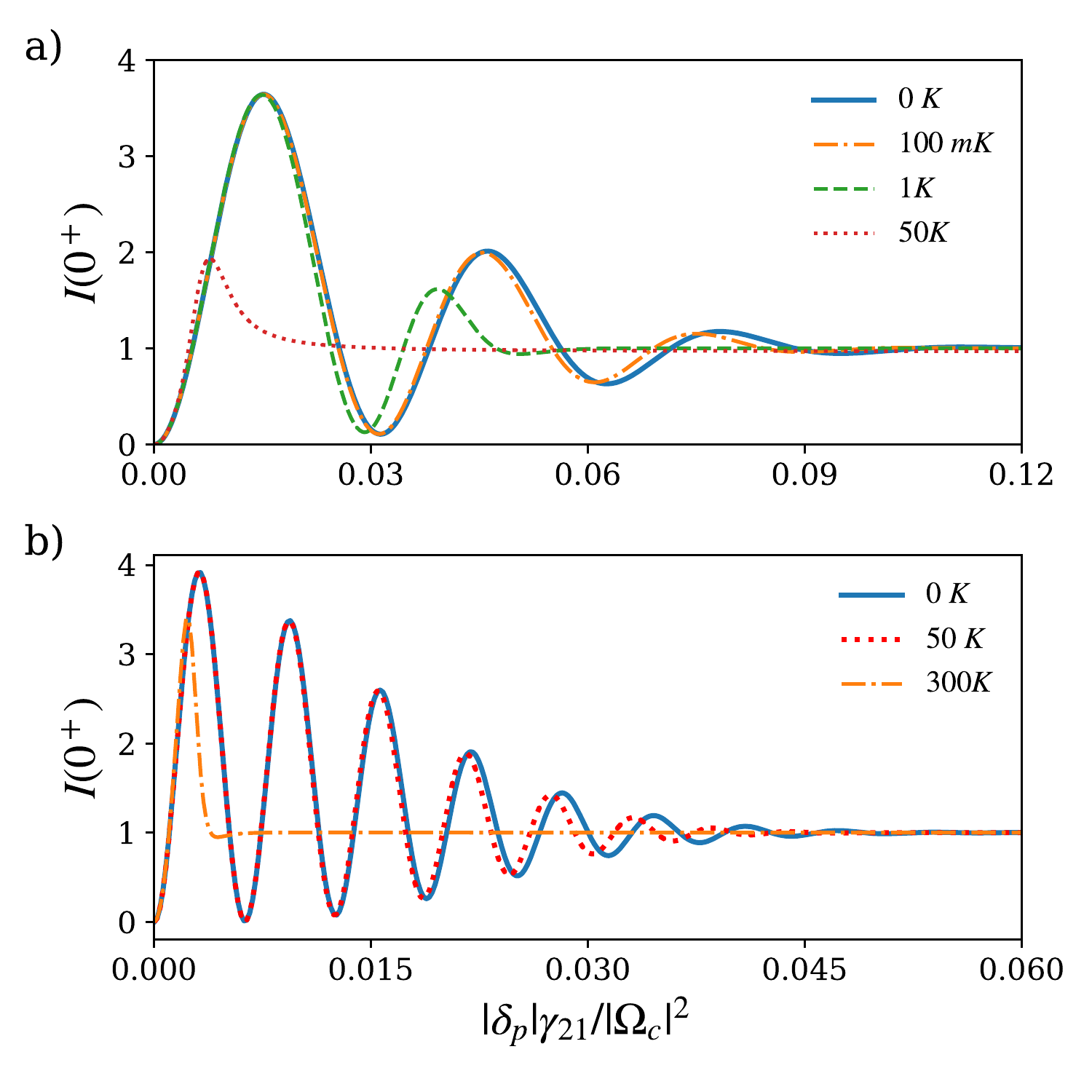}
    \caption{Effect of the finite temperature on the interference signal for $\gamma_{31}=0$. (a) The plots correspond to $b_0=200$ and temperatures of 0 K (blue curve), 100 mK (dash-dotted orange curve), 1 K (dashed green curve) and 50 K (dotted red curve). (b) The plots correspond to $b_0=1000$ and temperatures of 0 K (blue curve), 50 K (dotted red curve) and 300 K (dash-dotted orange curve). We use here the parameters for rubidium $D2$ line ($\gamma_{21}/2\pi=6\,$MHz, $k/2\pi=1.28\times 10^6\,$ m$^{\textrm{-1}}$)}
    \label{EffectOfTemp}
\end{figure}

If the temperature of the atomic ensemble is finite, we expect similar results as far as the cooperative characteristic rate $\tau_f^{-1}$ is larger than the Doppler broadening \cite{KCA19}, and the relaxation of the ground state coherence remains small. At room temperature, one can use, for example, a buffer gas~\cite{BNW97} or an anti-reflection coated vapor cell~\cite{BB66,GKR05, KHP11}. To investigate the effect of Doppler broadening, we plot in Fig. \ref{EffectOfTemp}a the interference fringes for $b_0=200$ at various temperatures for a rubidium ensemble excited on the $D2$ line ($\gamma_{21}/2\pi=6\,$MHz, and $k/2\pi=1.28\times 10^6\,$m$^{\textrm{-1}}$, where $k$ is the wave-number of the probe). Clearly at $b_0=200$, the interference fringes can be observed using cold atomic samples (with temperature, $T < 100$ mK) but are not observable in samples above $50\,$K. More precisely, Doppler broadening can be disregarded if the OD of the medium is such that $\tau_f^{-1} \gg k\bar{v}$, where $\bar{v}$ is the thermal velocity of the ensemble. To observe an interference signal at room temperature, a larger OD is required,  as shown in Fig. \ref{EffectOfTemp}b for $b_0=1000$.

\subsubsection*{Square-pulse amplitude modulation}

\begin{figure}[bt]
\includegraphics[scale=0.57]{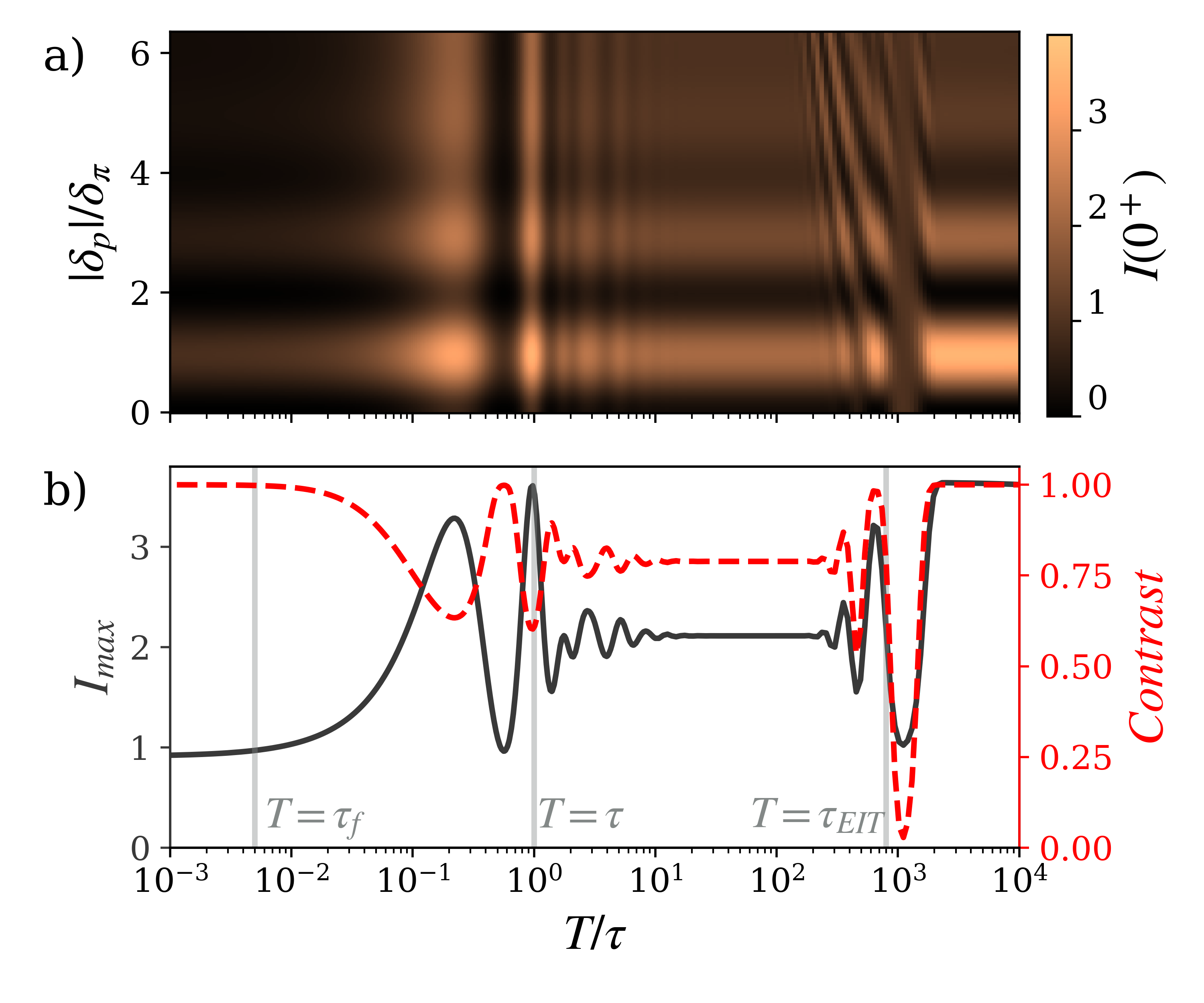}
\caption{\label{pulse} (a) $I(0^+)$ as functions of the pulse period $T/\tau$, and probe detuning $|\delta_p|/\delta_{\pi}$. (b) The first fringe maximum $I_\text{max}$ and the fringe contrast are plotted against the pulse period $T/\tau$. The parameters for the calculations are $b_0=200$, $\Omega_c = \gamma_{21}/2$ and $\gamma_{31}=0$. }
\end{figure}

We have so far analyzed the case of long probe duration where the steady-state regime of EIT transparency is achieved before the pulse extinction. Considering realistic experimental parameters for alkali metal atoms of $b_0=200$, $\gamma_{21}/2\pi=6$ MHz, and $\Omega_c=\gamma_{21}/2$, the square pulse duration has to be longer than $\tau_{EIT}\approx 21\,\mu$s.  The homodyne measurement takes place within the decay time of  the flash, typically $\tau_f\sim130$~ps, implying an unfavorably low duty cycle. We now show that it is possible to improve the duty cycle by several orders of magnitude. To this end, we consider a periodic square-pulse amplitude modulation of the probe beam, with
\begin{equation}
   {E}_i(t) = \begin{cases}
    1, \qquad -\frac{T}{2} < t < 0,\\
    0, \qquad  0 \geq t \leq \frac{T}{2},\\
    \end{cases}
\end{equation}
and ${E}_i(t+nT) = {E}_i(t)$ for all integers $n$.  In Fig.~\ref{pulse}a, we plot the transmitted intensity at the falling edge, $I(0^+)$, for $b_0=200$ and $\Omega_c = \gamma_{21}/2$, in the 2D parameter space of pulse period $T$ and probe detuning $\delta_p$. For the slow regime of $T\gg \tau_{EIT}$, the EIT steady-state is achieved within the probe duration. The interference fringe pattern is independent of the modulation period and depends only on $\Delta\phi$ as given in Eq.~(\ref{Dphi}). 

The same fringe pattern with varying contrast is also observed for $T\ll\tau_{EIT}$, suggesting that the self-homodyne detection technique can be extended to higher modulation frequencies. In this regime, as far as the EIT response of the medium is concerned, the probe beam can be considered as a time-averaged constant field with half the amplitude. Indeed, when $T\ll\tau_{EIT}$, the sidebands are all outside of the narrow EIT transparency window, leaving only the carrier component with a normalized amplitude of 0.5 inside the transparency window. Thus, we expect that $E_{EIT}(t)$ has a constant amplitude, with a phase shift of $\phi_{EIT}$. As the phase of $E_s(0)$ is essentially constant over the small $\delta_p$ range that we are considering, the fringe pattern is governed mainly by $\phi_{EIT}$. This leads to the same fringe pattern across a large range of modulation periods.

The performance of our square-pulse amplitude modulation technique is summarized in Fig.~\ref{pulse}b, by the black curve for the maximum flash $I_\text{max}$ and the red dashed curve for the fringe contrast $(I_\text{max}-I_\text{min})/(I_\text{max}+I_\text{min})$. $I_\text{max}$ is defined as the intensity of the first fringe maximum of $I(0^+)$. The minimum intensity, $I_\text{min}$, is found at resonance where the destructive interference between $E_s$ and $E_{EIT}$ gives an anti-flash dip at the falling edge of the square pulse. 

\begin{figure}[bt]
\includegraphics[scale=0.57]{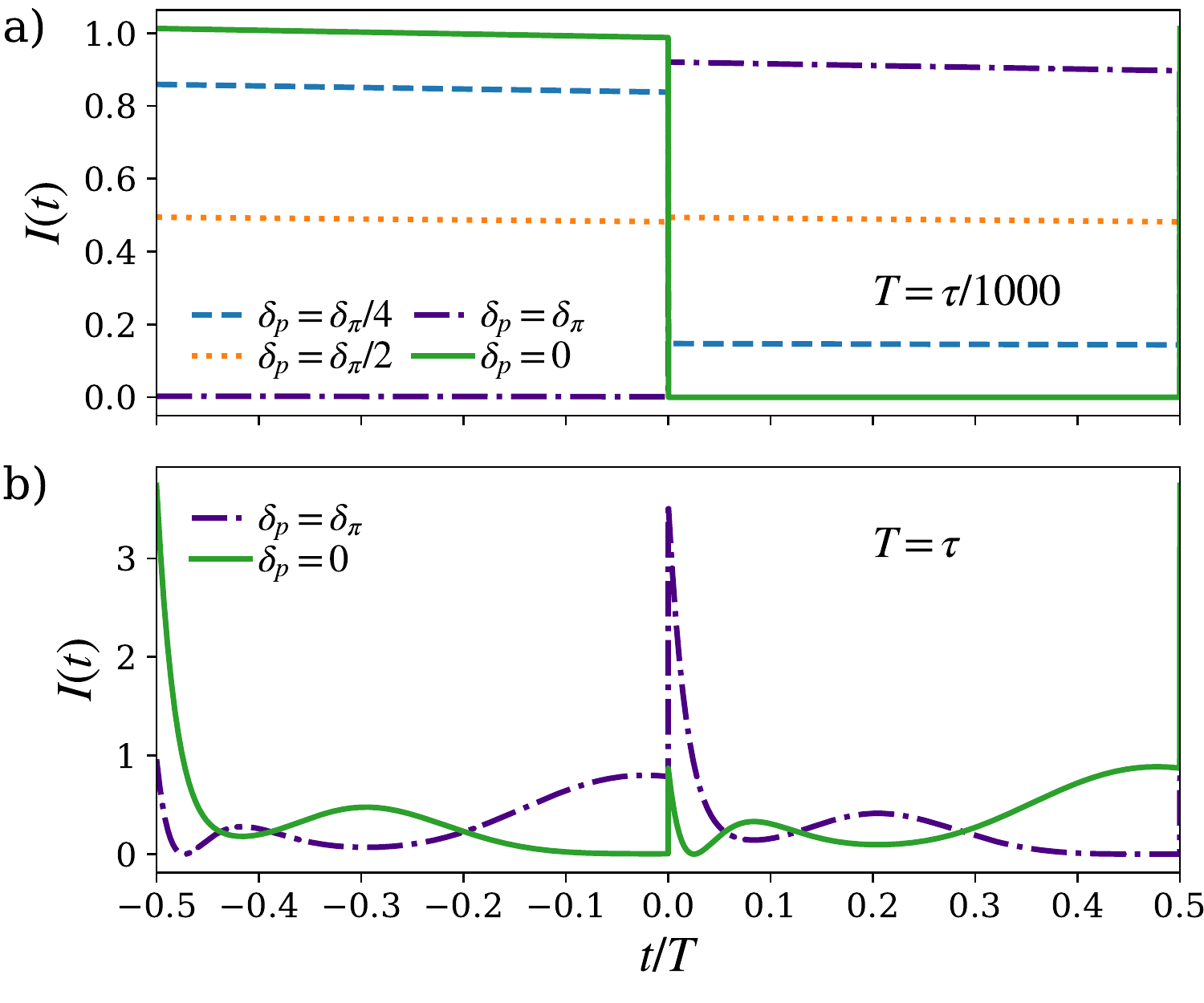}
\caption{\label{fig:fig4} The transmitted probe signal with the square-pulse amplitude modulation technique, for $b_0=200$, $\Omega_c = \gamma_{21}/2$ $\gamma_{31}=0$, (a) $T=\tau/1000\ (T<\tau_f$ regime) at various $\delta_p$ \textit{viz.,} $\delta_{\pi}/4$ (blue dashed line), $\delta_{\pi}/2$ (orange dotted), $\delta_{\pi}$ (violet dash-dotted) and at resonance (solid green) (b) $T=\tau\ (T>\tau_f$ regime) for $\delta_p = \delta_{\pi}$ (violet dash-dotted) and at resonance (solid green).}
\end{figure}

There are two regimes with unity contrast. The first one at $T\gg \tau_{EIT}$ is the steady-state regime discussed previously. The second regime at $T\ll\tau_f$ (\ie, $T/\tau\ll b_0^{-1}$) is of practical interest because of the much higher duty cycle. Due to the high modulation frequency, all the sidebands are far-detuned and transmitted unaffected through the atomic medium. The sidebands sum up to an amplitude of $E_i(t)-1/2$. Neglecting the off-resonance absorption in the transparency window, the transmitted carrier field is $\exp(i\phi_{EIT})/2$. Adding this to the sidebands, the total transmitted signal is $|E_i(t)+[\exp(i\phi_{EIT})-1]/2|^2$, and retains the square-pulse shape.  Some examples of the calculated transmitted signals at different $\delta_p$ values are shown in \fig{fig4}a. As expected, the transmitted signal shows a square-pulse profile without emitting any flashes, since the short modulation period does not leave sufficient time for the flashes to develop. We further observed a flip of the square pulse when $|\delta_p|$ goes from zero to $\delta_{\pi}$, consistent with the $\pi$ phase shift experienced by the transmitted carrier component. A slight reduction of the signal occurs off-resonance due to the residual absorption of the carrier component in the transparency window. This signal, sensitive to the two-photon detuning, can be easily detected using standard amplitude-modulation spectroscopy. The fields decomposition of Eq.~(\ref{E_threelevel}) can also be applied in this regime. Since $T\ll\tau_f\ll\tau_{EIT}$, both the two-level and EIT responses of the atomic medium experience effectively a ``time-averaged'' constant incident beam. As an example, we discuss the resonant case where  $E_s\approx -0.5$ and $E_{EIT}\approx 0.5$. Eq.~(\ref{E_threelevel}) gives $E(t)=E_i(t)$ and $I(t)\approx|E_i(t)|^2$\, as expected for the resonant case.  This reasoning can be easily extended to the case of $\delta_p\neq 0$. 

In the $T\ll\tau_f$ regime discussed above, we achieved a large improvement in the measurement duty cycle of the two-photon detuning, but the fringe maximum is smaller as compared to the steady-state regime because of the absence of flashes. To take advantage of the cooperative enhancement of the transient signals, we consider now the intermediate regime of $\tau_f<T\lesssim \tau$.  While the duty cycle is still high, the modulation frequency is now long enough for the flashes to develop at both the leading and falling edges. At certain modulation periods, this leads to a higher $I_\text{max}$ (see \fig{fig4}b for the case of $T\gamma_{21} = 1$). Here, the subsequent flash occurs during the transient of the previous flash. As a result, $I_\text{max}$ and the contrast oscillate as functions of the modulation period (see Fig.~\ref{pulse}b). This regime was previously studied in the two-level system~\cite{KYD15}. Importantly, the contrast  remains high for the cases where $I_\text{max}$ is maximal, making them useful for future applications. \\

\section{Conclusion}
In conclusion, we have studied a new approach to determine the two-photon resonance frequency of an optically thick three-level $\Lambda$-system, in the transient regime. This technique is based on a self-aligned homodyne detection of the slow light of EIT through its interference with a flash. The latter is the fast cooperative emission of light after an abrupt extinction of the incident field and has a time-scale that is well-separated from the EIT slow light. We showed that amplitude-modulation spectroscopy using square pulses allows a sensitive determination of the two-photon resonance, thanks to cooperative effects. This technique could be considered for CPT clocks~\cite{V05}, velocimetry~\cite{CLH20} and magnetometry~\cite{FMS00,BGK02} applications. However, the quantitative comparison of the performances of this technique with already established methods such as Ramsey spectroscopy \cite{THE82, PST08, ZGC05, barantsev2018line,EBI13,BRD15,LYT17} is beyond the scope of the paper and it is left for future works. Importantly, this technique  can be performed in a wide range of modulation frequencies that are higher than transition linewidth, relying on a \textit{time-averaged} regime of the EIT response. The optimum modulation frequency is a compromise between the fast and the intermediate regime. In the fast regime, the transmitted signal retains its square-pulse profile, with unity contrast in the fringe pattern.  The intermediate regime, on the other hand, benefits from the cooperative flash emission resulting in larger signal amplitude. At the same time, the fringe contrast remains high. \\

\section*{Acknowledgment}
The authors thank Thomas Zanon-Willette for critical reading of the manuscript. K.P. would like to acknowledge the funding from DST of grant No. DST/ICPS/QuST/Theme-3/2019.  This work was supported by the CQT/MoE funding Grant No. R-710-002-016-271.

\appendix
\addappheadtotoc
\counterwithin{figure}{section}
\section{}
\label{Appendix}

With a minor algebraic manipulation of Eq.~(4) in the main text, the susceptibility for a weak probe beam of detuning $\delta_p$ within the $\Lambda$-system is given by,
\begin{equation}
\label{chi}
\chi(\delta_p)=\frac{-6\pi \rho}{k^3} \frac{\frac{\gamma_{21}}{2}[\delta_p-\delta_c+\frac{i\gamma_{31}}{2}]}{(\delta_p+i\frac{\gamma_{21}}{2})(\delta_p-\delta_c+i\frac{\gamma_{31}}{2})-\frac{|\Omega_{c}|^2}{4}}\nonumber
\end{equation}

For $\delta_c=0$ , the above expression reads, 
\begin{equation}
\chi(\delta_p)=\frac{-6\pi \rho}{k^3} \frac{\frac{\gamma_{21}}{2}[\delta_p+\frac{i\gamma_{31}}{2}]}{\delta_p^2-\frac{\gamma_{21}\gamma_{31}}{4}-\frac{|\Omega_{c}|^2}{4}+i\delta_p\frac{\gamma_{21}+\gamma_{31}}{2}}\nonumber
\end{equation}
 
In the limit of $\gamma_{31}$ and $\delta_p \ll\Omega_c,\gamma_{21}$, we can write, $\delta_p^2-\frac{\gamma_{21}\gamma_{31}}{4}-\frac{|\Omega_{c}|^2}{4}+i\delta_p\frac{\gamma_{21}+\gamma_{31}}{2}\approx-\frac{|\Omega_{c}|^2}{4}+i(\delta_p+i\frac{\gamma_{31}}{2})\frac{\gamma_{21}}{2}.$ Hence, the expression for $\chi$ can be written as 
\begin{equation}
\chi(\delta_p)\approx\frac{-6\pi \rho}{k^3} \frac{\delta_p+\frac{i\gamma_{31}}{2}}{-\frac{|\Omega_{c}|^2}{2\gamma_{21}}-\frac{\gamma_{31}}{2}+i\delta_p}\nonumber
\end{equation}

Keeping to the first order in $\delta_p$ and $\gamma_{31}$, we have
\begin{equation}
\chi(\delta_p)\approx\frac{6\pi \rho}{k^3} \frac{2\gamma_{21}}{|\Omega_{c}|^2}\left(\delta_p+i\frac{\gamma_{31}}{2}\right)\nonumber
\end{equation}

The real and the imaginary part of $\chi$ can be written as follows:
\begin{align}
\text{Re}\{\chi(\delta_p)\}\approx\frac{6\pi \rho}{k^3}\frac{2\gamma_{21}}{|\Omega_c|^2}\delta_p,\nonumber\\
\text{Im}\{\chi(\delta_p)\}\approx\frac{6\pi \rho}{k^3}\frac{\gamma_{21}}{|\Omega_c|^2}\gamma_{31}\label{chiapprox}.
\end{align}

We now derive the time delay. The group velocity, $v_g$ the probe laser can be written as 
\begin{equation}
v_g=\frac{c_0}{n(\omega_p)+\frac{\partial n}{\partial\omega_p}\omega_p}\nonumber
\end{equation}
where $c_0$ is the speed of light in vacuum, and $n=1+\text{Re}(\chi)/2$ is the refractive index of the atomic medium.
Furthermore, we know that $n(\omega_p)\ll\frac{\partial n}{\partial \omega_p}\omega_p$, so $v_g\approx\frac{c_0}{\frac{\partial n}{\partial\omega_p}\omega_p}$.
Putting the value of the Re$\{\chi\}$ from Eq.~(\ref{chiapprox}) and using $k=\omega_p/c_0$, we get
\begin{equation}
n\approx1+\frac{1}{2}\frac{6\pi\rho c_0^3}{\omega_p^3}\frac{2\gamma_{21}}{|\Omega_c|^2}\delta_p
\end{equation}
where $\omega_0=\frac{E_2-E_1}{\hbar}$. Using $\delta_p=\omega_p-\omega_0$, we have
\begin{equation}
\frac{\partial n}{\partial \omega_p}\approx\frac{1}{2}\frac{6\pi\rho c_0^3}{\omega_p^3}\left[1-3\frac{\delta_p}{\omega_p}\right]\frac{2\gamma_{21}}{|\Omega_c|^2}.\nonumber
\end{equation}
For the quasi-resonant case where $\frac{\delta_p}{\omega_p}\ll1$, the above equation becomes,
\begin{equation}
\frac{\partial n}{\partial \omega_p}\approx\frac{1}{2}\frac{6\pi\rho c_0^3}{\omega_p^3}\frac{2\gamma_{21}}{|\Omega_c|^2},\nonumber
\end{equation}
and the group velocity reads,
\begin{equation}
v_g\approx\frac{\omega_p^2}{6\pi\rho c_0^2} \frac{|\Omega_c|^2}{\gamma_{21}}.
\end{equation}

Finally, the time delay defined as $\tau_{EIT}=L/v_g-L/c_0\approx L/v_g$ is given by, 
\begin{equation}
\tau_{EIT}\approx\frac{6\pi\rho c_0^2L}{\omega_p^2}\frac{\gamma_{21}}{|\Omega_c|^2}=\frac{b_0\gamma_{21}}{|\Omega_c|^2}.
\end{equation}

\bibliography{eitrefsallmodified}

\begin{thebibliography}{57}%
\makeatletter
\providecommand \@ifxundefined [1]{%
 \@ifx{#1\undefined}
}%
\providecommand \@ifnum [1]{%
 \ifnum #1\expandafter \@firstoftwo
 \else \expandafter \@secondoftwo
 \fi
}%
\providecommand \@ifx [1]{%
 \ifx #1\expandafter \@firstoftwo
 \else \expandafter \@secondoftwo
 \fi
}%
\providecommand \natexlab [1]{#1}%
\providecommand \enquote  [1]{``#1''}%
\providecommand \bibnamefont  [1]{#1}%
\providecommand \bibfnamefont [1]{#1}%
\providecommand \citenamefont [1]{#1}%
\providecommand \href@noop [0]{\@secondoftwo}%
\providecommand \href [0]{\begingroup \@sanitize@url \@href}%
\providecommand \@href[1]{\@@startlink{#1}\@@href}%
\providecommand \@@href[1]{\endgroup#1\@@endlink}%
\providecommand \@sanitize@url [0]{\catcode `\\12\catcode `\$12\catcode
  `\&12\catcode `\#12\catcode `\^12\catcode `\_12\catcode `\%12\relax}%
\providecommand \@@startlink[1]{}%
\providecommand \@@endlink[0]{}%
\providecommand \url  [0]{\begingroup\@sanitize@url \@url }%
\providecommand \@url [1]{\endgroup\@href {#1}{\urlprefix }}%
\providecommand \urlprefix  [0]{URL }%
\providecommand \Eprint [0]{\href }%
\providecommand \doibase [0]{https://doi.org/}%
\providecommand \selectlanguage [0]{\@gobble}%
\providecommand \bibinfo  [0]{\@secondoftwo}%
\providecommand \bibfield  [0]{\@secondoftwo}%
\providecommand \translation [1]{[#1]}%
\providecommand \BibitemOpen [0]{}%
\providecommand \bibitemStop [0]{}%
\providecommand \bibitemNoStop [0]{.\EOS\space}%
\providecommand \EOS [0]{\spacefactor3000\relax}%
\providecommand \BibitemShut  [1]{\csname bibitem#1\endcsname}%
\let\auto@bib@innerbib\@empty
\bibitem [{\citenamefont {Alzetta}\ \emph {et~al.}(1976)\citenamefont
  {Alzetta}, \citenamefont {Gozzini}, \citenamefont {Moi},\ and\ \citenamefont
  {Orriols}}]{alzetta1976experimental}%
  \BibitemOpen
  \bibfield  {author} {\bibinfo {author} {\bibfnamefont {G.}~\bibnamefont
  {Alzetta}}, \bibinfo {author} {\bibfnamefont {A.}~\bibnamefont {Gozzini}},
  \bibinfo {author} {\bibfnamefont {L.}~\bibnamefont {Moi}},\ and\ \bibinfo
  {author} {\bibfnamefont {G.}~\bibnamefont {Orriols}},\ }\bibfield  {title}
  {\bibinfo {title} {An experimental method for the observation of rf
  transitions and laser beat resonances in oriented na vapour},\ }\href@noop {}
  {\bibfield  {journal} {\bibinfo  {journal} {Il Nuovo Cimento B (1971-1996)}\
  }\textbf {\bibinfo {volume} {36}},\ \bibinfo {pages} {5} (\bibinfo {year}
  {1976})}\BibitemShut {NoStop}%
\bibitem [{\citenamefont {Boller}\ \emph {et~al.}(1991)\citenamefont {Boller},
  \citenamefont {Imamo{\u{g}}lu},\ and\ \citenamefont
  {Harris}}]{boller1991observation}%
  \BibitemOpen
  \bibfield  {author} {\bibinfo {author} {\bibfnamefont {K.-J.}\ \bibnamefont
  {Boller}}, \bibinfo {author} {\bibfnamefont {A.}~\bibnamefont
  {Imamo{\u{g}}lu}},\ and\ \bibinfo {author} {\bibfnamefont {S.~E.}\
  \bibnamefont {Harris}},\ }\bibfield  {title} {\bibinfo {title} {Observation
  of electromagnetically induced transparency},\ }\href@noop {} {\bibfield
  {journal} {\bibinfo  {journal} {Physical Review Letters}\ }\textbf {\bibinfo
  {volume} {66}},\ \bibinfo {pages} {2593} (\bibinfo {year}
  {1991})}\BibitemShut {NoStop}%
\bibitem [{\citenamefont {Arimondo}\ and\ \citenamefont
  {Orriols}(1976)}]{arimondo1976nonabsorbing}%
  \BibitemOpen
  \bibfield  {author} {\bibinfo {author} {\bibfnamefont {E.}~\bibnamefont
  {Arimondo}}\ and\ \bibinfo {author} {\bibfnamefont {G.}~\bibnamefont
  {Orriols}},\ }\bibfield  {title} {\bibinfo {title} {Nonabsorbing atomic
  coherences by coherent two-photon transitions in a three-level optical
  pumping},\ }\href@noop {} {\bibfield  {journal} {\bibinfo  {journal} {Lettere
  al Nuovo Cimento (1971-1985)}\ }\textbf {\bibinfo {volume} {17}},\ \bibinfo
  {pages} {333} (\bibinfo {year} {1976})}\BibitemShut {NoStop}%
\bibitem [{\citenamefont {Hau}\ \emph {et~al.}(1999)\citenamefont {Hau},
  \citenamefont {Harris}, \citenamefont {Dutton},\ and\ \citenamefont
  {Behroozi}}]{HHD99}%
  \BibitemOpen
  \bibfield  {author} {\bibinfo {author} {\bibfnamefont {L.~V.}\ \bibnamefont
  {Hau}}, \bibinfo {author} {\bibfnamefont {S.~E.}\ \bibnamefont {Harris}},
  \bibinfo {author} {\bibfnamefont {Z.}~\bibnamefont {Dutton}},\ and\ \bibinfo
  {author} {\bibfnamefont {C.~H.}\ \bibnamefont {Behroozi}},\ }\bibfield
  {title} {\bibinfo {title} {Light speed reduction to 17 metres per second in
  an ultracold atomic gas},\ }\href {https://doi.org/10.1038/17561} {\bibfield
  {journal} {\bibinfo  {journal} {Nature}\ }\textbf {\bibinfo {volume} {397}},\
  \bibinfo {pages} {594} (\bibinfo {year} {1999})}\BibitemShut {NoStop}%
\bibitem [{\citenamefont {Katz}\ and\ \citenamefont
  {Firstenberg}(2018)}]{KAF18}%
  \BibitemOpen
  \bibfield  {author} {\bibinfo {author} {\bibfnamefont {O.}~\bibnamefont
  {Katz}}\ and\ \bibinfo {author} {\bibfnamefont {O.}~\bibnamefont
  {Firstenberg}},\ }\bibfield  {title} {\bibinfo {title} {Light storage for one
  second in room-temperature alkali vapor},\ }\href@noop {} {\bibfield
  {journal} {\bibinfo  {journal} {Nature Communications}\ }\textbf {\bibinfo
  {volume} {9}},\ \bibinfo {pages} {2074} (\bibinfo {year} {2018})}\BibitemShut
  {NoStop}%
\bibitem [{\citenamefont {Bussi{\`e}res}\ \emph {et~al.}(2013)\citenamefont
  {Bussi{\`e}res}, \citenamefont {Sangouard}, \citenamefont {Afzelius},
  \citenamefont {de~Riedmatten}, \citenamefont {Simon},\ and\ \citenamefont
  {Tittel}}]{BSA13}%
  \BibitemOpen
  \bibfield  {author} {\bibinfo {author} {\bibfnamefont {F.}~\bibnamefont
  {Bussi{\`e}res}}, \bibinfo {author} {\bibfnamefont {N.}~\bibnamefont
  {Sangouard}}, \bibinfo {author} {\bibfnamefont {M.}~\bibnamefont {Afzelius}},
  \bibinfo {author} {\bibfnamefont {H.}~\bibnamefont {de~Riedmatten}}, \bibinfo
  {author} {\bibfnamefont {C.}~\bibnamefont {Simon}},\ and\ \bibinfo {author}
  {\bibfnamefont {W.}~\bibnamefont {Tittel}},\ }\bibfield  {title} {\bibinfo
  {title} {Prospective applications of optical quantum memories},\ }\href
  {https://doi.org/10.1080/09500340.2013.856482} {\bibfield  {journal}
  {\bibinfo  {journal} {Journal of Modern Optics}\ }\textbf {\bibinfo {volume}
  {60}},\ \bibinfo {pages} {1519} (\bibinfo {year} {2013})},\ \Eprint
  {https://arxiv.org/abs/https://doi.org/10.1080/09500340.2013.856482}
  {https://doi.org/10.1080/09500340.2013.856482} \BibitemShut {NoStop}%
\bibitem [{\citenamefont {Hsiao}\ \emph {et~al.}(2018)\citenamefont {Hsiao},
  \citenamefont {Tsai}, \citenamefont {Chen}, \citenamefont {Lin},
  \citenamefont {Hung}, \citenamefont {Lee}, \citenamefont {Chen},
  \citenamefont {Chen}, \citenamefont {Yu},\ and\ \citenamefont
  {Chen}}]{HTC18}%
  \BibitemOpen
  \bibfield  {author} {\bibinfo {author} {\bibfnamefont {Y.-F.}\ \bibnamefont
  {Hsiao}}, \bibinfo {author} {\bibfnamefont {P.-J.}\ \bibnamefont {Tsai}},
  \bibinfo {author} {\bibfnamefont {H.-S.}\ \bibnamefont {Chen}}, \bibinfo
  {author} {\bibfnamefont {S.-X.}\ \bibnamefont {Lin}}, \bibinfo {author}
  {\bibfnamefont {C.-C.}\ \bibnamefont {Hung}}, \bibinfo {author}
  {\bibfnamefont {C.-H.}\ \bibnamefont {Lee}}, \bibinfo {author} {\bibfnamefont
  {Y.-H.}\ \bibnamefont {Chen}}, \bibinfo {author} {\bibfnamefont {Y.-F.}\
  \bibnamefont {Chen}}, \bibinfo {author} {\bibfnamefont {I.~A.}\ \bibnamefont
  {Yu}},\ and\ \bibinfo {author} {\bibfnamefont {Y.-C.}\ \bibnamefont {Chen}},\
  }\bibfield  {title} {\bibinfo {title} {Highly efficient coherent optical
  memory based on electromagnetically induced transparency},\ }\href
  {https://doi.org/10.1103/PhysRevLett.120.183602} {\bibfield  {journal}
  {\bibinfo  {journal} {Phys. Rev. Lett.}\ }\textbf {\bibinfo {volume} {120}},\
  \bibinfo {pages} {183602} (\bibinfo {year} {2018})}\BibitemShut {NoStop}%
\bibitem [{\citenamefont {Hafiz}\ \emph {et~al.}(2016)\citenamefont {Hafiz},
  \citenamefont {Liu}, \citenamefont {Gu{\'{e}}randel}, \citenamefont
  {Clercq},\ and\ \citenamefont {Boudot}}]{MXT16}%
  \BibitemOpen
  \bibfield  {author} {\bibinfo {author} {\bibfnamefont {M.~A.}\ \bibnamefont
  {Hafiz}}, \bibinfo {author} {\bibfnamefont {X.}~\bibnamefont {Liu}}, \bibinfo
  {author} {\bibfnamefont {S.}~\bibnamefont {Gu{\'{e}}randel}}, \bibinfo
  {author} {\bibfnamefont {E.~D.}\ \bibnamefont {Clercq}},\ and\ \bibinfo
  {author} {\bibfnamefont {R.}~\bibnamefont {Boudot}},\ }\bibfield  {title}
  {\bibinfo {title} {A {CPT}-based cs vapor cell atomic clock with a short-term
  fractional frequency stability of 3 x 10-13$\uptau$-1/2},\ }\href
  {https://doi.org/10.1088/1742-6596/723/1/012013} {\bibfield  {journal}
  {\bibinfo  {journal} {Journal of Physics: Conference Series}\ }\textbf
  {\bibinfo {volume} {723}},\ \bibinfo {pages} {012013} (\bibinfo {year}
  {2016})}\BibitemShut {NoStop}%
\bibitem [{\citenamefont {Vanier}(2005)}]{V05}%
  \BibitemOpen
  \bibfield  {author} {\bibinfo {author} {\bibfnamefont {J.}~\bibnamefont
  {Vanier}},\ }\bibfield  {title} {\bibinfo {title} {Atomic clocks based on
  coherent population trapping: a review},\ }\href
  {https://doi.org/10.1007/s00340-005-1905-3} {\bibfield  {journal} {\bibinfo
  {journal} {Applied Physics B}\ }\textbf {\bibinfo {volume} {81}},\ \bibinfo
  {pages} {421} (\bibinfo {year} {2005})}\BibitemShut {NoStop}%
\bibitem [{\citenamefont {Liu}\ \emph {et~al.}(2017{\natexlab{a}})\citenamefont
  {Liu}, \citenamefont {Ivanov}, \citenamefont {Yudin}, \citenamefont
  {Kitching},\ and\ \citenamefont {Donley}}]{LIY17}%
  \BibitemOpen
  \bibfield  {author} {\bibinfo {author} {\bibfnamefont {X.}~\bibnamefont
  {Liu}}, \bibinfo {author} {\bibfnamefont {E.}~\bibnamefont {Ivanov}},
  \bibinfo {author} {\bibfnamefont {V.~I.}\ \bibnamefont {Yudin}}, \bibinfo
  {author} {\bibfnamefont {J.}~\bibnamefont {Kitching}},\ and\ \bibinfo
  {author} {\bibfnamefont {E.~A.}\ \bibnamefont {Donley}},\ }\bibfield  {title}
  {\bibinfo {title} {Low-drift coherent population trapping clock based on
  laser-cooled atoms and high-coherence excitation fields},\ }\href
  {https://doi.org/10.1103/PhysRevApplied.8.054001} {\bibfield  {journal}
  {\bibinfo  {journal} {Phys. Rev. Applied}\ }\textbf {\bibinfo {volume} {8}},\
  \bibinfo {pages} {054001} (\bibinfo {year} {2017}{\natexlab{a}})}\BibitemShut
  {NoStop}%
\bibitem [{\citenamefont {Liu}\ \emph {et~al.}(2013)\citenamefont {Liu},
  \citenamefont {M\'erolla}, \citenamefont {Gu\'erandel}, \citenamefont
  {Gorecki}, \citenamefont {de~Clercq},\ and\ \citenamefont {Boudot}}]{LMJ13}%
  \BibitemOpen
  \bibfield  {author} {\bibinfo {author} {\bibfnamefont {X.}~\bibnamefont
  {Liu}}, \bibinfo {author} {\bibfnamefont {J.-M.}\ \bibnamefont {M\'erolla}},
  \bibinfo {author} {\bibfnamefont {S.}~\bibnamefont {Gu\'erandel}}, \bibinfo
  {author} {\bibfnamefont {C.}~\bibnamefont {Gorecki}}, \bibinfo {author}
  {\bibfnamefont {E.}~\bibnamefont {de~Clercq}},\ and\ \bibinfo {author}
  {\bibfnamefont {R.}~\bibnamefont {Boudot}},\ }\bibfield  {title} {\bibinfo
  {title} {Coherent-population-trapping resonances in buffer-gas-filled
  cs-vapor cells with push-pull optical pumping},\ }\href
  {https://doi.org/10.1103/PhysRevA.87.013416} {\bibfield  {journal} {\bibinfo
  {journal} {Phys. Rev. A}\ }\textbf {\bibinfo {volume} {87}},\ \bibinfo
  {pages} {013416} (\bibinfo {year} {2013})}\BibitemShut {NoStop}%
\bibitem [{\citenamefont {Liu}\ \emph {et~al.}(2017{\natexlab{b}})\citenamefont
  {Liu}, \citenamefont {Yudin}, \citenamefont {Taichenachev}, \citenamefont
  {Kitching},\ and\ \citenamefont {Donley}}]{LYT17}%
  \BibitemOpen
  \bibfield  {author} {\bibinfo {author} {\bibfnamefont {X.}~\bibnamefont
  {Liu}}, \bibinfo {author} {\bibfnamefont {V.~I.}\ \bibnamefont {Yudin}},
  \bibinfo {author} {\bibfnamefont {A.~V.}\ \bibnamefont {Taichenachev}},
  \bibinfo {author} {\bibfnamefont {J.}~\bibnamefont {Kitching}},\ and\
  \bibinfo {author} {\bibfnamefont {E.~A.}\ \bibnamefont {Donley}},\ }\bibfield
   {title} {\bibinfo {title} {High contrast dark resonances in a cold-atom
  clock probed with counterpropagating circularly polarized beams},\ }\href
  {https://doi.org/10.1063/1.5001179} {\bibfield  {journal} {\bibinfo
  {journal} {Applied Physics Letters}\ }\textbf {\bibinfo {volume} {111}},\
  \bibinfo {pages} {224102} (\bibinfo {year} {2017}{\natexlab{b}})},\ \Eprint
  {https://arxiv.org/abs/https://doi.org/10.1063/1.5001179}
  {https://doi.org/10.1063/1.5001179} \BibitemShut {NoStop}%
\bibitem [{\citenamefont {Fang}\ \emph {et~al.}(2020)\citenamefont {Fang},
  \citenamefont {Han}, \citenamefont {Jiang}, \citenamefont {Qiu},
  \citenamefont {Guo}, \citenamefont {Zhao}, \citenamefont {Huang},
  \citenamefont {Lu},\ and\ \citenamefont {Lee}}]{RCX20}%
  \BibitemOpen
  \bibfield  {author} {\bibinfo {author} {\bibfnamefont {R.}~\bibnamefont
  {Fang}}, \bibinfo {author} {\bibfnamefont {C.}~\bibnamefont {Han}}, \bibinfo
  {author} {\bibfnamefont {X.}~\bibnamefont {Jiang}}, \bibinfo {author}
  {\bibfnamefont {Y.}~\bibnamefont {Qiu}}, \bibinfo {author} {\bibfnamefont
  {Y.}~\bibnamefont {Guo}}, \bibinfo {author} {\bibfnamefont {M.}~\bibnamefont
  {Zhao}}, \bibinfo {author} {\bibfnamefont {J.}~\bibnamefont {Huang}},
  \bibinfo {author} {\bibfnamefont {B.}~\bibnamefont {Lu}},\ and\ \bibinfo
  {author} {\bibfnamefont {C.}~\bibnamefont {Lee}},\ }\bibfield  {title}
  {\bibinfo {title} {Temporal spinwave fabry-perot interferometry via coherent
  population trapping},\ }\href {https://arxiv.org/abs/2008.12562} {\bibfield
  {journal} {\bibinfo  {journal} {arXiv:2008.12562}\ } (\bibinfo {year}
  {2020})}\BibitemShut {NoStop}%
\bibitem [{\citenamefont {Sedlacek}\ \emph {et~al.}(2012)\citenamefont
  {Sedlacek}, \citenamefont {Schwettmann}, \citenamefont {Kubler},
  \citenamefont {Low}, \citenamefont {Pfau},\ and\ \citenamefont
  {Shaffer}}]{JSA12}%
  \BibitemOpen
  \bibfield  {author} {\bibinfo {author} {\bibfnamefont {J.~A.}\ \bibnamefont
  {Sedlacek}}, \bibinfo {author} {\bibfnamefont {A.}~\bibnamefont
  {Schwettmann}}, \bibinfo {author} {\bibfnamefont {H.}~\bibnamefont {Kubler}},
  \bibinfo {author} {\bibfnamefont {R.}~\bibnamefont {Low}}, \bibinfo {author}
  {\bibfnamefont {T.}~\bibnamefont {Pfau}},\ and\ \bibinfo {author}
  {\bibfnamefont {J.~P.}\ \bibnamefont {Shaffer}},\ }\bibfield  {title}
  {\bibinfo {title} {Microwave electrometry with rydberg atoms in a vapour cell
  using bright atomic resonances},\ }\href
  {https://doi.org/doi:10.1038/nphys2423} {\bibfield  {journal} {\bibinfo
  {journal} {Nature physics}\ }\textbf {\bibinfo {volume} {8}},\ \bibinfo
  {pages} {819} (\bibinfo {year} {2012})}\BibitemShut {NoStop}%
\bibitem [{\citenamefont {Shylla}\ \emph {et~al.}(2018)\citenamefont {Shylla},
  \citenamefont {Nyakang'o},\ and\ \citenamefont {Pandey}}]{SNP18}%
  \BibitemOpen
  \bibfield  {author} {\bibinfo {author} {\bibfnamefont {D.}~\bibnamefont
  {Shylla}}, \bibinfo {author} {\bibfnamefont {E.~O.}\ \bibnamefont
  {Nyakang'o}},\ and\ \bibinfo {author} {\bibfnamefont {K.}~\bibnamefont
  {Pandey}},\ }\bibfield  {title} {\bibinfo {title} {Highly sensitive atomic
  based mw interferometry},\ }\href
  {https://doi.org/10.1038/s41598-018-27011-1} {\bibfield  {journal} {\bibinfo
  {journal} {Scientific Reports}\ }\textbf {\bibinfo {volume} {8}},\ \bibinfo
  {pages} {8692} (\bibinfo {year} {2018})}\BibitemShut {NoStop}%
\bibitem [{\citenamefont {Wade}\ \emph {et~al.}(2018)\citenamefont {Wade},
  \citenamefont {Marcuzzi}, \citenamefont {Levi}, \citenamefont {Kondo},
  \citenamefont {Lesanovsky}, \citenamefont {Adams},\ and\ \citenamefont
  {Weatherill}}]{WML18}%
  \BibitemOpen
  \bibfield  {author} {\bibinfo {author} {\bibfnamefont {C.~G.}\ \bibnamefont
  {Wade}}, \bibinfo {author} {\bibfnamefont {M.}~\bibnamefont {Marcuzzi}},
  \bibinfo {author} {\bibfnamefont {E.}~\bibnamefont {Levi}}, \bibinfo {author}
  {\bibfnamefont {J.~M.}\ \bibnamefont {Kondo}}, \bibinfo {author}
  {\bibfnamefont {I.}~\bibnamefont {Lesanovsky}}, \bibinfo {author}
  {\bibfnamefont {C.~S.}\ \bibnamefont {Adams}},\ and\ \bibinfo {author}
  {\bibfnamefont {K.~J.}\ \bibnamefont {Weatherill}},\ }\bibfield  {title}
  {\bibinfo {title} {A terahertz-driven non-equilibrium phase transition in a
  room temperature atomic vapour},\ }\href
  {https://doi.org/10.1038/s41467-018-05597-4} {\bibfield  {journal} {\bibinfo
  {journal} {Nature Communications}\ }\textbf {\bibinfo {volume} {9}},\
  \bibinfo {pages} {3567} (\bibinfo {year} {2018})}\BibitemShut {NoStop}%
\bibitem [{\citenamefont {Lam}\ \emph {et~al.}(2021)\citenamefont {Lam},
  \citenamefont {Pal}, \citenamefont {Vogt}, \citenamefont {Kiffner},\ and\
  \citenamefont {Li}}]{lam2021directional}%
  \BibitemOpen
  \bibfield  {author} {\bibinfo {author} {\bibfnamefont {M.}~\bibnamefont
  {Lam}}, \bibinfo {author} {\bibfnamefont {S.~B.}\ \bibnamefont {Pal}},
  \bibinfo {author} {\bibfnamefont {T.}~\bibnamefont {Vogt}}, \bibinfo {author}
  {\bibfnamefont {M.}~\bibnamefont {Kiffner}},\ and\ \bibinfo {author}
  {\bibfnamefont {W.}~\bibnamefont {Li}},\ }\bibfield  {title} {\bibinfo
  {title} {Directional thz generation in hot $\textrm{Rb}$ vapor excited to a
  rydberg state},\ }\href@noop {} {\bibfield  {journal} {\bibinfo  {journal}
  {Optics Letters}\ }\textbf {\bibinfo {volume} {46}},\ \bibinfo {pages} {1017}
  (\bibinfo {year} {2021})}\BibitemShut {NoStop}%
\bibitem [{\citenamefont {Aspect}\ \emph {et~al.}(1988)\citenamefont {Aspect},
  \citenamefont {Arimondo}, \citenamefont {Kaiser}, \citenamefont
  {Vansteenkiste},\ and\ \citenamefont {Cohen-Tannoudji}}]{aspect1988laser}%
  \BibitemOpen
  \bibfield  {author} {\bibinfo {author} {\bibfnamefont {A.}~\bibnamefont
  {Aspect}}, \bibinfo {author} {\bibfnamefont {E.}~\bibnamefont {Arimondo}},
  \bibinfo {author} {\bibfnamefont {R.}~\bibnamefont {Kaiser}}, \bibinfo
  {author} {\bibfnamefont {N.}~\bibnamefont {Vansteenkiste}},\ and\ \bibinfo
  {author} {\bibfnamefont {C.}~\bibnamefont {Cohen-Tannoudji}},\ }\bibfield
  {title} {\bibinfo {title} {Laser cooling below the one-photon recoil energy
  by velocity-selective coherent population trapping},\ }\href@noop {}
  {\bibfield  {journal} {\bibinfo  {journal} {Physical Review Letters}\
  }\textbf {\bibinfo {volume} {61}},\ \bibinfo {pages} {826} (\bibinfo {year}
  {1988})}\BibitemShut {NoStop}%
\bibitem [{\citenamefont {Wilkowski}\ \emph {et~al.}(2009)\citenamefont
  {Wilkowski}, \citenamefont {Chalony}, \citenamefont {Kaiser},\ and\
  \citenamefont {Kastberg}}]{Wilkowski_2009}%
  \BibitemOpen
  \bibfield  {author} {\bibinfo {author} {\bibfnamefont {D.}~\bibnamefont
  {Wilkowski}}, \bibinfo {author} {\bibfnamefont {M.}~\bibnamefont {Chalony}},
  \bibinfo {author} {\bibfnamefont {R.}~\bibnamefont {Kaiser}},\ and\ \bibinfo
  {author} {\bibfnamefont {A.}~\bibnamefont {Kastberg}},\ }\bibfield  {title}
  {\bibinfo {title} {Low- and high-intensity velocity selective coherent
  population trapping in a two-level system},\ }\href
  {https://doi.org/10.1209/0295-5075/86/53001} {\bibfield  {journal} {\bibinfo
  {journal} {{EPL} (Europhysics Letters)}\ }\textbf {\bibinfo {volume} {86}},\
  \bibinfo {pages} {53001} (\bibinfo {year} {2009})}\BibitemShut {NoStop}%
\bibitem [{\citenamefont {Chen}\ \emph {et~al.}(2020)\citenamefont {Chen},
  \citenamefont {Lim}, \citenamefont {Huang}, \citenamefont {Dumke},\ and\
  \citenamefont {Lan}}]{CLH20}%
  \BibitemOpen
  \bibfield  {author} {\bibinfo {author} {\bibfnamefont {Z.}~\bibnamefont
  {Chen}}, \bibinfo {author} {\bibfnamefont {H.~M.}\ \bibnamefont {Lim}},
  \bibinfo {author} {\bibfnamefont {C.}~\bibnamefont {Huang}}, \bibinfo
  {author} {\bibfnamefont {R.}~\bibnamefont {Dumke}},\ and\ \bibinfo {author}
  {\bibfnamefont {S.-Y.}\ \bibnamefont {Lan}},\ }\bibfield  {title} {\bibinfo
  {title} {Quantum-enhanced velocimetry with doppler-broadened atomic vapor},\
  }\href {https://doi.org/10.1103/PhysRevLett.124.093202} {\bibfield  {journal}
  {\bibinfo  {journal} {Phys. Rev. Lett.}\ }\textbf {\bibinfo {volume} {124}},\
  \bibinfo {pages} {093202} (\bibinfo {year} {2020})}\BibitemShut {NoStop}%
\bibitem [{\citenamefont {Fleischhauer}\ \emph {et~al.}(2005)\citenamefont
  {Fleischhauer}, \citenamefont {Imamoglu},\ and\ \citenamefont
  {Marangos}}]{FIM05}%
  \BibitemOpen
  \bibfield  {author} {\bibinfo {author} {\bibfnamefont {M.}~\bibnamefont
  {Fleischhauer}}, \bibinfo {author} {\bibfnamefont {A.}~\bibnamefont
  {Imamoglu}},\ and\ \bibinfo {author} {\bibfnamefont {J.~P.}\ \bibnamefont
  {Marangos}},\ }\bibfield  {title} {\bibinfo {title} {Electromagnetically
  induced transparency: Optics in coherent media},\ }\href
  {https://doi.org/10.1103/RevModPhys.77.633} {\bibfield  {journal} {\bibinfo
  {journal} {Rev. Mod. Phys.}\ }\textbf {\bibinfo {volume} {77}},\ \bibinfo
  {pages} {633} (\bibinfo {year} {2005})}\BibitemShut {NoStop}%
\bibitem [{\citenamefont {Tebben}\ \emph {et~al.}(2021)\citenamefont {Tebben},
  \citenamefont {Hainaut}, \citenamefont {Salzinger}, \citenamefont {Geier},
  \citenamefont {Franz}, \citenamefont {Pohl}, \citenamefont {G\"arttner},
  \citenamefont {Z\"urn},\ and\ \citenamefont
  {Weidem\"uller}}]{tebben2021nonlinear}%
  \BibitemOpen
  \bibfield  {author} {\bibinfo {author} {\bibfnamefont {A.}~\bibnamefont
  {Tebben}}, \bibinfo {author} {\bibfnamefont {C.}~\bibnamefont {Hainaut}},
  \bibinfo {author} {\bibfnamefont {A.}~\bibnamefont {Salzinger}}, \bibinfo
  {author} {\bibfnamefont {S.}~\bibnamefont {Geier}}, \bibinfo {author}
  {\bibfnamefont {T.}~\bibnamefont {Franz}}, \bibinfo {author} {\bibfnamefont
  {T.}~\bibnamefont {Pohl}}, \bibinfo {author} {\bibfnamefont {M.}~\bibnamefont
  {G\"arttner}}, \bibinfo {author} {\bibfnamefont {G.}~\bibnamefont {Z\"urn}},\
  and\ \bibinfo {author} {\bibfnamefont {M.}~\bibnamefont {Weidem\"uller}},\
  }\bibfield  {title} {\bibinfo {title} {Nonlinear absorption in interacting
  rydberg electromagnetically-induced-transparency spectra on two-photon
  resonance},\ }\href {https://doi.org/10.1103/PhysRevA.103.063710} {\bibfield
  {journal} {\bibinfo  {journal} {Phys. Rev. A}\ }\textbf {\bibinfo {volume}
  {103}},\ \bibinfo {pages} {063710} (\bibinfo {year} {2021})}\BibitemShut
  {NoStop}%
\bibitem [{\citenamefont {Firstenberg}\ \emph {et~al.}(2013)\citenamefont
  {Firstenberg}, \citenamefont {Peyronel}, \citenamefont {Liang}, \citenamefont
  {Gorshkov}, \citenamefont {Lukin},\ and\ \citenamefont
  {Vuleti{\'c}}}]{firstenberg2013attractive}%
  \BibitemOpen
  \bibfield  {author} {\bibinfo {author} {\bibfnamefont {O.}~\bibnamefont
  {Firstenberg}}, \bibinfo {author} {\bibfnamefont {T.}~\bibnamefont
  {Peyronel}}, \bibinfo {author} {\bibfnamefont {Q.-Y.}\ \bibnamefont {Liang}},
  \bibinfo {author} {\bibfnamefont {A.~V.}\ \bibnamefont {Gorshkov}}, \bibinfo
  {author} {\bibfnamefont {M.~D.}\ \bibnamefont {Lukin}},\ and\ \bibinfo
  {author} {\bibfnamefont {V.}~\bibnamefont {Vuleti{\'c}}},\ }\bibfield
  {title} {\bibinfo {title} {Attractive photons in a quantum nonlinear
  medium},\ }\href@noop {} {\bibfield  {journal} {\bibinfo  {journal} {Nature}\
  }\textbf {\bibinfo {volume} {502}},\ \bibinfo {pages} {71} (\bibinfo {year}
  {2013})}\BibitemShut {NoStop}%
\bibitem [{\citenamefont {Carusotto}\ and\ \citenamefont
  {Ciuti}(2013)}]{carusotto2013quantum}%
  \BibitemOpen
  \bibfield  {author} {\bibinfo {author} {\bibfnamefont {I.}~\bibnamefont
  {Carusotto}}\ and\ \bibinfo {author} {\bibfnamefont {C.}~\bibnamefont
  {Ciuti}},\ }\bibfield  {title} {\bibinfo {title} {Quantum fluids of light},\
  }\href@noop {} {\bibfield  {journal} {\bibinfo  {journal} {Reviews of Modern
  Physics}\ }\textbf {\bibinfo {volume} {85}},\ \bibinfo {pages} {299}
  (\bibinfo {year} {2013})}\BibitemShut {NoStop}%
\bibitem [{\citenamefont {Collombon}\ \emph {et~al.}(2019)\citenamefont
  {Collombon}, \citenamefont {Chatou}, \citenamefont {Hagel}, \citenamefont
  {Pedregosa-Gutierrez}, \citenamefont {Houssin}, \citenamefont {Knoop},\ and\
  \citenamefont {Champenois}}]{PhysRevApplied.12.034035}%
  \BibitemOpen
  \bibfield  {author} {\bibinfo {author} {\bibfnamefont {M.}~\bibnamefont
  {Collombon}}, \bibinfo {author} {\bibfnamefont {C.}~\bibnamefont {Chatou}},
  \bibinfo {author} {\bibfnamefont {G.}~\bibnamefont {Hagel}}, \bibinfo
  {author} {\bibfnamefont {J.}~\bibnamefont {Pedregosa-Gutierrez}}, \bibinfo
  {author} {\bibfnamefont {M.}~\bibnamefont {Houssin}}, \bibinfo {author}
  {\bibfnamefont {M.}~\bibnamefont {Knoop}},\ and\ \bibinfo {author}
  {\bibfnamefont {C.}~\bibnamefont {Champenois}},\ }\bibfield  {title}
  {\bibinfo {title} {Experimental demonstration of three-photon coherent
  population trapping in an ion cloud},\ }\href
  {https://doi.org/10.1103/PhysRevApplied.12.034035} {\bibfield  {journal}
  {\bibinfo  {journal} {Phys. Rev. Applied}\ }\textbf {\bibinfo {volume}
  {12}},\ \bibinfo {pages} {034035} (\bibinfo {year} {2019})}\BibitemShut
  {NoStop}%
\bibitem [{\citenamefont {Lukin}\ \emph {et~al.}(1997)\citenamefont {Lukin},
  \citenamefont {Fleischhauer}, \citenamefont {Zibrov}, \citenamefont
  {Robinson}, \citenamefont {Velichansky}, \citenamefont {Hollberg},\ and\
  \citenamefont {Scully}}]{LFZ97}%
  \BibitemOpen
  \bibfield  {author} {\bibinfo {author} {\bibfnamefont {M.~D.}\ \bibnamefont
  {Lukin}}, \bibinfo {author} {\bibfnamefont {M.}~\bibnamefont {Fleischhauer}},
  \bibinfo {author} {\bibfnamefont {A.~S.}\ \bibnamefont {Zibrov}}, \bibinfo
  {author} {\bibfnamefont {H.~G.}\ \bibnamefont {Robinson}}, \bibinfo {author}
  {\bibfnamefont {V.~L.}\ \bibnamefont {Velichansky}}, \bibinfo {author}
  {\bibfnamefont {L.}~\bibnamefont {Hollberg}},\ and\ \bibinfo {author}
  {\bibfnamefont {M.~O.}\ \bibnamefont {Scully}},\ }\bibfield  {title}
  {\bibinfo {title} {Spectroscopy in dense coherent media: Line narrowing and
  interference effects},\ }\href {https://doi.org/10.1103/PhysRevLett.79.2959}
  {\bibfield  {journal} {\bibinfo  {journal} {Phys. Rev. Lett.}\ }\textbf
  {\bibinfo {volume} {79}},\ \bibinfo {pages} {2959} (\bibinfo {year}
  {1997})}\BibitemShut {NoStop}%
\bibitem [{\citenamefont {Godone}\ \emph {et~al.}(2002)\citenamefont {Godone},
  \citenamefont {Levi}, \citenamefont {Micalizio},\ and\ \citenamefont
  {Vanier}}]{GLM02}%
  \BibitemOpen
  \bibfield  {author} {\bibinfo {author} {\bibfnamefont {A.}~\bibnamefont
  {Godone}}, \bibinfo {author} {\bibfnamefont {F.}~\bibnamefont {Levi}},
  \bibinfo {author} {\bibfnamefont {S.}~\bibnamefont {Micalizio}},\ and\
  \bibinfo {author} {\bibfnamefont {J.}~\bibnamefont {Vanier}},\ }\bibfield
  {title} {\bibinfo {title} {Dark-line in optically-thick vapors: inversion
  phenomena and line width narrowing},\ }\href
  {https://doi.org/10.1140/e10053-002-0001-z} {\bibfield  {journal} {\bibinfo
  {journal} {Eur. Phys. J. D}\ }\textbf {\bibinfo {volume} {18}},\ \bibinfo
  {pages} {5} (\bibinfo {year} {2002})}\BibitemShut {NoStop}%
\bibitem [{\citenamefont {Bentley}\ \emph {et~al.}(2000)\citenamefont
  {Bentley}, \citenamefont {Liu},\ and\ \citenamefont {Liao}}]{BLL00}%
  \BibitemOpen
  \bibfield  {author} {\bibinfo {author} {\bibfnamefont {C.~L.}\ \bibnamefont
  {Bentley}}, \bibinfo {author} {\bibfnamefont {J.}~\bibnamefont {Liu}},\ and\
  \bibinfo {author} {\bibfnamefont {Y.}~\bibnamefont {Liao}},\ }\bibfield
  {title} {\bibinfo {title} {Cavity electromagnetically induced transparency of
  driven-three-level atoms: A transparent window narrowing below a natural
  width},\ }\href {https://doi.org/10.1103/PhysRevA.61.023811} {\bibfield
  {journal} {\bibinfo  {journal} {Phys. Rev. A}\ }\textbf {\bibinfo {volume}
  {61}},\ \bibinfo {pages} {023811} (\bibinfo {year} {2000})}\BibitemShut
  {NoStop}%
\bibitem [{\citenamefont {Wang}\ \emph {et~al.}(2000)\citenamefont {Wang},
  \citenamefont {Goorskey}, \citenamefont {Burkett},\ and\ \citenamefont
  {Xiao}}]{WGB00}%
  \BibitemOpen
  \bibfield  {author} {\bibinfo {author} {\bibfnamefont {H.}~\bibnamefont
  {Wang}}, \bibinfo {author} {\bibfnamefont {D.~J.}\ \bibnamefont {Goorskey}},
  \bibinfo {author} {\bibfnamefont {W.~H.}\ \bibnamefont {Burkett}},\ and\
  \bibinfo {author} {\bibfnamefont {M.}~\bibnamefont {Xiao}},\ }\bibfield
  {title} {\bibinfo {title} {Cavity-linewidth narrowing by means of
  electromagnetically induced transparency},\ }\href
  {https://doi.org/10.1364/OL.25.001732} {\bibfield  {journal} {\bibinfo
  {journal} {Opt. Lett.}\ }\textbf {\bibinfo {volume} {25}},\ \bibinfo {pages}
  {1732} (\bibinfo {year} {2000})}\BibitemShut {NoStop}%
\bibitem [{\citenamefont {Lukin}\ \emph {et~al.}(1998)\citenamefont {Lukin},
  \citenamefont {Fleischhauer}, \citenamefont {Scully},\ and\ \citenamefont
  {Velichansky}}]{LFS98}%
  \BibitemOpen
  \bibfield  {author} {\bibinfo {author} {\bibfnamefont {M.~D.}\ \bibnamefont
  {Lukin}}, \bibinfo {author} {\bibfnamefont {M.}~\bibnamefont {Fleischhauer}},
  \bibinfo {author} {\bibfnamefont {M.~O.}\ \bibnamefont {Scully}},\ and\
  \bibinfo {author} {\bibfnamefont {V.~L.}\ \bibnamefont {Velichansky}},\
  }\bibfield  {title} {\bibinfo {title} {Intracavity electromagnetically
  induced transparency},\ }\href {https://doi.org/10.1364/OL.23.000295}
  {\bibfield  {journal} {\bibinfo  {journal} {Opt. Lett.}\ }\textbf {\bibinfo
  {volume} {23}},\ \bibinfo {pages} {295} (\bibinfo {year} {1998})}\BibitemShut
  {NoStop}%
\bibitem [{\citenamefont {Hernandez}\ \emph {et~al.}(2007)\citenamefont
  {Hernandez}, \citenamefont {Zhang},\ and\ \citenamefont {Zhu}}]{HZZ07}%
  \BibitemOpen
  \bibfield  {author} {\bibinfo {author} {\bibfnamefont {G.}~\bibnamefont
  {Hernandez}}, \bibinfo {author} {\bibfnamefont {J.}~\bibnamefont {Zhang}},\
  and\ \bibinfo {author} {\bibfnamefont {Y.}~\bibnamefont {Zhu}},\ }\bibfield
  {title} {\bibinfo {title} {Vacuum rabi splitting and intracavity dark state
  in a cavity-atom system},\ }\href
  {https://doi.org/10.1103/PhysRevA.76.053814} {\bibfield  {journal} {\bibinfo
  {journal} {Phys. Rev. A}\ }\textbf {\bibinfo {volume} {76}},\ \bibinfo
  {pages} {053814} (\bibinfo {year} {2007})}\BibitemShut {NoStop}%
\bibitem [{\citenamefont {Wu}\ \emph {et~al.}(2008)\citenamefont {Wu},
  \citenamefont {Gea-Banacloche},\ and\ \citenamefont {Xiao}}]{WGX08}%
  \BibitemOpen
  \bibfield  {author} {\bibinfo {author} {\bibfnamefont {H.}~\bibnamefont
  {Wu}}, \bibinfo {author} {\bibfnamefont {J.}~\bibnamefont {Gea-Banacloche}},\
  and\ \bibinfo {author} {\bibfnamefont {M.}~\bibnamefont {Xiao}},\ }\bibfield
  {title} {\bibinfo {title} {Observation of intracavity electromagnetically
  induced transparency and polariton resonances in a doppler-broadened
  medium},\ }\href {https://doi.org/10.1103/PhysRevLett.100.173602} {\bibfield
  {journal} {\bibinfo  {journal} {Phys. Rev. Lett.}\ }\textbf {\bibinfo
  {volume} {100}},\ \bibinfo {pages} {173602} (\bibinfo {year}
  {2008})}\BibitemShut {NoStop}%
\bibitem [{\citenamefont {Thomas}\ \emph {et~al.}(1982)\citenamefont {Thomas},
  \citenamefont {Hemmer}, \citenamefont {Ezekiel}, \citenamefont {Leiby},
  \citenamefont {Picard},\ and\ \citenamefont {Willis}}]{THE82}%
  \BibitemOpen
  \bibfield  {author} {\bibinfo {author} {\bibfnamefont {J.~E.}\ \bibnamefont
  {Thomas}}, \bibinfo {author} {\bibfnamefont {P.~R.}\ \bibnamefont {Hemmer}},
  \bibinfo {author} {\bibfnamefont {S.}~\bibnamefont {Ezekiel}}, \bibinfo
  {author} {\bibfnamefont {C.~C.}\ \bibnamefont {Leiby}}, \bibinfo {author}
  {\bibfnamefont {R.~H.}\ \bibnamefont {Picard}},\ and\ \bibinfo {author}
  {\bibfnamefont {C.~R.}\ \bibnamefont {Willis}},\ }\bibfield  {title}
  {\bibinfo {title} {Observation of ramsey fringes using a stimulated,
  resonance raman transition in a sodium atomic beam},\ }\href
  {https://doi.org/10.1103/PhysRevLett.48.867} {\bibfield  {journal} {\bibinfo
  {journal} {Phys. Rev. Lett.}\ }\textbf {\bibinfo {volume} {48}},\ \bibinfo
  {pages} {867} (\bibinfo {year} {1982})}\BibitemShut {NoStop}%
\bibitem [{\citenamefont {Pati}\ \emph {et~al.}(2008)\citenamefont {Pati},
  \citenamefont {Salit}, \citenamefont {Tripathi},\ and\ \citenamefont
  {Shahriar}}]{PST08}%
  \BibitemOpen
  \bibfield  {author} {\bibinfo {author} {\bibfnamefont {G.}~\bibnamefont
  {Pati}}, \bibinfo {author} {\bibfnamefont {K.}~\bibnamefont {Salit}},
  \bibinfo {author} {\bibfnamefont {R.}~\bibnamefont {Tripathi}},\ and\
  \bibinfo {author} {\bibfnamefont {M.}~\bibnamefont {Shahriar}},\ }\bibfield
  {title} {\bibinfo {title} {Demonstration of raman ramsey fringes using time
  delayed optical pulses in rubidium vapor},\ }\href
  {https://doi.org/10.1016/j.optcom.2008.05.056} {\bibfield  {journal}
  {\bibinfo  {journal} {Optics Communications}\ }\textbf {\bibinfo {volume}
  {281}},\ \bibinfo {pages} {4676} (\bibinfo {year} {2008})}\BibitemShut
  {NoStop}%
\bibitem [{\citenamefont {Zanon}\ \emph {et~al.}(2005)\citenamefont {Zanon},
  \citenamefont {Guerandel}, \citenamefont {de~Clercq}, \citenamefont
  {Holleville}, \citenamefont {Dimarcq},\ and\ \citenamefont
  {Clairon}}]{ZGC05}%
  \BibitemOpen
  \bibfield  {author} {\bibinfo {author} {\bibfnamefont {T.}~\bibnamefont
  {Zanon}}, \bibinfo {author} {\bibfnamefont {S.}~\bibnamefont {Guerandel}},
  \bibinfo {author} {\bibfnamefont {E.}~\bibnamefont {de~Clercq}}, \bibinfo
  {author} {\bibfnamefont {D.}~\bibnamefont {Holleville}}, \bibinfo {author}
  {\bibfnamefont {N.}~\bibnamefont {Dimarcq}},\ and\ \bibinfo {author}
  {\bibfnamefont {A.}~\bibnamefont {Clairon}},\ }\bibfield  {title} {\bibinfo
  {title} {High contrast ramsey fringes with coherent-population-trapping
  pulses in a double lambda atomic system},\ }\href
  {https://doi.org/10.1103/PhysRevLett.94.193002} {\bibfield  {journal}
  {\bibinfo  {journal} {Phys. Rev. Lett.}\ }\textbf {\bibinfo {volume} {94}},\
  \bibinfo {pages} {193002} (\bibinfo {year} {2005})}\BibitemShut {NoStop}%
\bibitem [{\citenamefont {Barantsev}\ \emph {et~al.}(2018)\citenamefont
  {Barantsev}, \citenamefont {Popov},\ and\ \citenamefont
  {Litvinov}}]{barantsev2018line}%
  \BibitemOpen
  \bibfield  {author} {\bibinfo {author} {\bibfnamefont {K.~A.}\ \bibnamefont
  {Barantsev}}, \bibinfo {author} {\bibfnamefont {E.~N.}\ \bibnamefont
  {Popov}},\ and\ \bibinfo {author} {\bibfnamefont {A.~N.}\ \bibnamefont
  {Litvinov}},\ }\bibfield  {title} {\bibinfo {title} {Line shape of coherent
  population trapping resonance in the $\lambda$-scheme under ramsey-type
  interrogation in an optically dense medium},\ }\href@noop {} {\bibfield
  {journal} {\bibinfo  {journal} {Quantum electronics}\ }\textbf {\bibinfo
  {volume} {48}},\ \bibinfo {pages} {615} (\bibinfo {year} {2018})}\BibitemShut
  {NoStop}%
\bibitem [{\citenamefont {Esnault}\ \emph {et~al.}(2013)\citenamefont
  {Esnault}, \citenamefont {Blanshan}, \citenamefont {Ivanov}, \citenamefont
  {Scholten}, \citenamefont {Kitching},\ and\ \citenamefont {Donley}}]{EBI13}%
  \BibitemOpen
  \bibfield  {author} {\bibinfo {author} {\bibfnamefont {F.-X.}\ \bibnamefont
  {Esnault}}, \bibinfo {author} {\bibfnamefont {E.}~\bibnamefont {Blanshan}},
  \bibinfo {author} {\bibfnamefont {E.~N.}\ \bibnamefont {Ivanov}}, \bibinfo
  {author} {\bibfnamefont {R.~E.}\ \bibnamefont {Scholten}}, \bibinfo {author}
  {\bibfnamefont {J.}~\bibnamefont {Kitching}},\ and\ \bibinfo {author}
  {\bibfnamefont {E.~A.}\ \bibnamefont {Donley}},\ }\bibfield  {title}
  {\bibinfo {title} {Cold-atom double-$\ensuremath{\Lambda}$ coherent
  population trapping clock},\ }\href
  {https://doi.org/10.1103/PhysRevA.88.042120} {\bibfield  {journal} {\bibinfo
  {journal} {Phys. Rev. A}\ }\textbf {\bibinfo {volume} {88}},\ \bibinfo
  {pages} {042120} (\bibinfo {year} {2013})}\BibitemShut {NoStop}%
\bibitem [{\citenamefont {Blanshan}\ \emph {et~al.}(2015)\citenamefont
  {Blanshan}, \citenamefont {Rochester}, \citenamefont {Donley},\ and\
  \citenamefont {Kitching}}]{BRD15}%
  \BibitemOpen
  \bibfield  {author} {\bibinfo {author} {\bibfnamefont {E.}~\bibnamefont
  {Blanshan}}, \bibinfo {author} {\bibfnamefont {S.~M.}\ \bibnamefont
  {Rochester}}, \bibinfo {author} {\bibfnamefont {E.~A.}\ \bibnamefont
  {Donley}},\ and\ \bibinfo {author} {\bibfnamefont {J.}~\bibnamefont
  {Kitching}},\ }\bibfield  {title} {\bibinfo {title} {Light shifts in a pulsed
  cold-atom coherent-population-trapping clock},\ }\href
  {https://doi.org/10.1103/PhysRevA.91.041401} {\bibfield  {journal} {\bibinfo
  {journal} {Phys. Rev. A}\ }\textbf {\bibinfo {volume} {91}},\ \bibinfo
  {pages} {041401} (\bibinfo {year} {2015})}\BibitemShut {NoStop}%
\bibitem [{\citenamefont {Kwong}\ \emph {et~al.}(2014)\citenamefont {Kwong},
  \citenamefont {Yang}, \citenamefont {Pramod}, \citenamefont {Pandey},
  \citenamefont {Delande}, \citenamefont {Pierrat},\ and\ \citenamefont
  {Wilkowski}}]{KYP14}%
  \BibitemOpen
  \bibfield  {author} {\bibinfo {author} {\bibfnamefont {C.~C.}\ \bibnamefont
  {Kwong}}, \bibinfo {author} {\bibfnamefont {T.}~\bibnamefont {Yang}},
  \bibinfo {author} {\bibfnamefont {M.~S.}\ \bibnamefont {Pramod}}, \bibinfo
  {author} {\bibfnamefont {K.}~\bibnamefont {Pandey}}, \bibinfo {author}
  {\bibfnamefont {D.}~\bibnamefont {Delande}}, \bibinfo {author} {\bibfnamefont
  {R.}~\bibnamefont {Pierrat}},\ and\ \bibinfo {author} {\bibfnamefont
  {D.}~\bibnamefont {Wilkowski}},\ }\bibfield  {title} {\bibinfo {title}
  {Cooperative emission of a coherent superflash of light},\ }\href
  {https://doi.org/10.1103/PhysRevLett.113.223601} {\bibfield  {journal}
  {\bibinfo  {journal} {Phys. Rev. Lett.}\ }\textbf {\bibinfo {volume} {113}},\
  \bibinfo {pages} {223601} (\bibinfo {year} {2014})}\BibitemShut {NoStop}%
\bibitem [{\citenamefont {Wei}\ \emph {et~al.}(2009)\citenamefont {Wei},
  \citenamefont {Chen}, \citenamefont {Loy}, \citenamefont {Wong},\ and\
  \citenamefont {Du}}]{WCL09}%
  \BibitemOpen
  \bibfield  {author} {\bibinfo {author} {\bibfnamefont {D.}~\bibnamefont
  {Wei}}, \bibinfo {author} {\bibfnamefont {J.~F.}\ \bibnamefont {Chen}},
  \bibinfo {author} {\bibfnamefont {M.~M.~T.}\ \bibnamefont {Loy}}, \bibinfo
  {author} {\bibfnamefont {G.~K.~L.}\ \bibnamefont {Wong}},\ and\ \bibinfo
  {author} {\bibfnamefont {S.}~\bibnamefont {Du}},\ }\bibfield  {title}
  {\bibinfo {title} {Optical precursors with electromagnetically induced
  transparency in cold atoms},\ }\href
  {https://doi.org/10.1103/PhysRevLett.103.093602} {\bibfield  {journal}
  {\bibinfo  {journal} {Phys. Rev. Lett.}\ }\textbf {\bibinfo {volume} {103}},\
  \bibinfo {pages} {093602} (\bibinfo {year} {2009})}\BibitemShut {NoStop}%
\bibitem [{\citenamefont {Datsyuk}\ \emph {et~al.}(2006)\citenamefont
  {Datsyuk}, \citenamefont {Sokolov}, \citenamefont {Kupriyanov},\ and\
  \citenamefont {Havey}}]{datsyuk2006diffuse}%
  \BibitemOpen
  \bibfield  {author} {\bibinfo {author} {\bibfnamefont {V.~M.}\ \bibnamefont
  {Datsyuk}}, \bibinfo {author} {\bibfnamefont {I.~M.}\ \bibnamefont
  {Sokolov}}, \bibinfo {author} {\bibfnamefont {D.~V.}\ \bibnamefont
  {Kupriyanov}},\ and\ \bibinfo {author} {\bibfnamefont {M.~D.}\ \bibnamefont
  {Havey}},\ }\bibfield  {title} {\bibinfo {title} {Diffuse light scattering
  dynamics under conditions of electromagnetically induced transparency},\
  }\href {https://doi.org/10.1103/PhysRevA.74.043812} {\bibfield  {journal}
  {\bibinfo  {journal} {Phys. Rev. A}\ }\textbf {\bibinfo {volume} {74}},\
  \bibinfo {pages} {043812} (\bibinfo {year} {2006})}\BibitemShut {NoStop}%
\bibitem [{\citenamefont {Sommerfeld}(1914)}]{SOM14}%
  \BibitemOpen
  \bibfield  {author} {\bibinfo {author} {\bibfnamefont {A.}~\bibnamefont
  {Sommerfeld}},\ }\bibfield  {title} {\bibinfo {title} {Uber die fortpflanzung
  des lichtes in disperdierenden medien},\ }\href@noop {} {\bibfield  {journal}
  {\bibinfo  {journal} {Ann. Phys. (Leipzig)}\ }\textbf {\bibinfo {volume}
  {44}} (\bibinfo {year} {1914})}\BibitemShut {NoStop}%
\bibitem [{\citenamefont {Brillouin}(1914)}]{BRI14}%
  \BibitemOpen
  \bibfield  {author} {\bibinfo {author} {\bibfnamefont {L.}~\bibnamefont
  {Brillouin}},\ }\bibfield  {title} {\bibinfo {title} {Uber die fortpflanzung
  des licht in disperdierenden medien},\ }\href@noop {} {\bibfield  {journal}
  {\bibinfo  {journal} {Ann. Phys. (Leipzig)}\ }\textbf {\bibinfo {volume}
  {44}} (\bibinfo {year} {1914})}\BibitemShut {NoStop}%
\bibitem [{\citenamefont {Bao}\ \emph {et~al.}(2014)\citenamefont {Bao},
  \citenamefont {Fang}, \citenamefont {Yang}, \citenamefont {Cui},\ and\
  \citenamefont {Wu}}]{BFY14}%
  \BibitemOpen
  \bibfield  {author} {\bibinfo {author} {\bibfnamefont {Q.-Q.}\ \bibnamefont
  {Bao}}, \bibinfo {author} {\bibfnamefont {B.}~\bibnamefont {Fang}}, \bibinfo
  {author} {\bibfnamefont {X.}~\bibnamefont {Yang}}, \bibinfo {author}
  {\bibfnamefont {C.-L.}\ \bibnamefont {Cui}},\ and\ \bibinfo {author}
  {\bibfnamefont {J.-H.}\ \bibnamefont {Wu}},\ }\bibfield  {title} {\bibinfo
  {title} {Marking slow light signals with fast optical precursors in the
  regime of electromagnetically induced transparency},\ }\href
  {https://doi.org/10.1364/JOSAB.31.000062} {\bibfield  {journal} {\bibinfo
  {journal} {J. Opt. Soc. Am. B}\ }\textbf {\bibinfo {volume} {31}},\ \bibinfo
  {pages} {62} (\bibinfo {year} {2014})}\BibitemShut {NoStop}%
\bibitem [{\citenamefont {Peng}\ \emph {et~al.}(2012)\citenamefont {Peng},
  \citenamefont {Yang}, \citenamefont {Chen}, \citenamefont {Xu},\ and\
  \citenamefont {Zhang}}]{PYC12}%
  \BibitemOpen
  \bibfield  {author} {\bibinfo {author} {\bibfnamefont {Y.~D.}\ \bibnamefont
  {Peng}}, \bibinfo {author} {\bibfnamefont {A.~H.}\ \bibnamefont {Yang}},
  \bibinfo {author} {\bibfnamefont {B.}~\bibnamefont {Chen}}, \bibinfo {author}
  {\bibfnamefont {Y.}~\bibnamefont {Xu}},\ and\ \bibinfo {author}
  {\bibfnamefont {L.~Y.}\ \bibnamefont {Zhang}},\ }\bibfield  {title} {\bibinfo
  {title} {Magnetically induced separation and enhancement of optical
  precursors via electromagnetically induced transparency},\ }\href
  {https://doi.org/10.1140/epjd/e2012-30488-2} {\bibfield  {journal} {\bibinfo
  {journal} {The European Physical Journal D}\ }\textbf {\bibinfo {volume}
  {66}},\ \bibinfo {pages} {296} (\bibinfo {year} {2012})}\BibitemShut
  {NoStop}%
\bibitem [{\citenamefont {Bloembergen}\ \emph {et~al.}(1948)\citenamefont
  {Bloembergen}, \citenamefont {Purcell},\ and\ \citenamefont {Pound}}]{BPP48}%
  \BibitemOpen
  \bibfield  {author} {\bibinfo {author} {\bibfnamefont {N.}~\bibnamefont
  {Bloembergen}}, \bibinfo {author} {\bibfnamefont {E.~M.}\ \bibnamefont
  {Purcell}},\ and\ \bibinfo {author} {\bibfnamefont {R.~V.}\ \bibnamefont
  {Pound}},\ }\bibfield  {title} {\bibinfo {title} {Relaxation effects in
  nuclear magnetic resonance absorption},\ }\href
  {https://doi.org/10.1103/PhysRev.73.679} {\bibfield  {journal} {\bibinfo
  {journal} {Phys. Rev.}\ }\textbf {\bibinfo {volume} {73}},\ \bibinfo {pages}
  {679} (\bibinfo {year} {1948})}\BibitemShut {NoStop}%
\bibitem [{\citenamefont {Toyoda}\ \emph {et~al.}(1997)\citenamefont {Toyoda},
  \citenamefont {Takahashi}, \citenamefont {Ishikawa},\ and\ \citenamefont
  {Yabuzaki}}]{TTI97}%
  \BibitemOpen
  \bibfield  {author} {\bibinfo {author} {\bibfnamefont {K.}~\bibnamefont
  {Toyoda}}, \bibinfo {author} {\bibfnamefont {Y.}~\bibnamefont {Takahashi}},
  \bibinfo {author} {\bibfnamefont {K.}~\bibnamefont {Ishikawa}},\ and\
  \bibinfo {author} {\bibfnamefont {T.}~\bibnamefont {Yabuzaki}},\ }\bibfield
  {title} {\bibinfo {title} {Optical free-induction decay of laser-cooled
  $^{85}\textrm{Rb}$},\ }\href {https://doi.org/10.1103/PhysRevA.56.1564}
  {\bibfield  {journal} {\bibinfo  {journal} {Phys. Rev. A}\ }\textbf {\bibinfo
  {volume} {56}},\ \bibinfo {pages} {1564} (\bibinfo {year}
  {1997})}\BibitemShut {NoStop}%
\bibitem [{\citenamefont {Kwong}\ \emph {et~al.}(2015)\citenamefont {Kwong},
  \citenamefont {Yang}, \citenamefont {Delande}, \citenamefont {Pierrat},\ and\
  \citenamefont {Wilkowski}}]{KYD15}%
  \BibitemOpen
  \bibfield  {author} {\bibinfo {author} {\bibfnamefont {C.~C.}\ \bibnamefont
  {Kwong}}, \bibinfo {author} {\bibfnamefont {T.}~\bibnamefont {Yang}},
  \bibinfo {author} {\bibfnamefont {D.}~\bibnamefont {Delande}}, \bibinfo
  {author} {\bibfnamefont {R.}~\bibnamefont {Pierrat}},\ and\ \bibinfo {author}
  {\bibfnamefont {D.}~\bibnamefont {Wilkowski}},\ }\bibfield  {title} {\bibinfo
  {title} {Cooperative emission of a pulse train in an optically thick
  scattering medium},\ }\href {https://doi.org/10.1103/PhysRevLett.115.223601}
  {\bibfield  {journal} {\bibinfo  {journal} {Phys. Rev. Lett.}\ }\textbf
  {\bibinfo {volume} {115}},\ \bibinfo {pages} {223601} (\bibinfo {year}
  {2015})}\BibitemShut {NoStop}%
\bibitem [{\citenamefont {Chalony}\ \emph {et~al.}(2011)\citenamefont
  {Chalony}, \citenamefont {Pierrat}, \citenamefont {Delande},\ and\
  \citenamefont {Wilkowski}}]{CPD11}%
  \BibitemOpen
  \bibfield  {author} {\bibinfo {author} {\bibfnamefont {M.}~\bibnamefont
  {Chalony}}, \bibinfo {author} {\bibfnamefont {R.}~\bibnamefont {Pierrat}},
  \bibinfo {author} {\bibfnamefont {D.}~\bibnamefont {Delande}},\ and\ \bibinfo
  {author} {\bibfnamefont {D.}~\bibnamefont {Wilkowski}},\ }\bibfield  {title}
  {\bibinfo {title} {Coherent flash of light emitted by a cold atomic cloud},\
  }\href {https://doi.org/10.1103/PhysRevA.84.011401} {\bibfield  {journal}
  {\bibinfo  {journal} {Phys. Rev. A}\ }\textbf {\bibinfo {volume} {84}},\
  \bibinfo {pages} {011401} (\bibinfo {year} {2011})}\BibitemShut {NoStop}%
\bibitem [{\citenamefont {Jackson}(1999)}]{J99}%
  \BibitemOpen
  \bibfield  {author} {\bibinfo {author} {\bibfnamefont {J.~D.}\ \bibnamefont
  {Jackson}},\ }\href@noop {} {\emph {\bibinfo {title} {Classical
  electrodynamics}}},\ \bibinfo {edition} {3rd}\ ed.\ (\bibinfo  {publisher}
  {Wiley},\ \bibinfo {address} {New York},\ \bibinfo {year} {1999})\BibitemShut
  {NoStop}%
\bibitem [{\citenamefont {Kwong}\ \emph {et~al.}(2019)\citenamefont {Kwong},
  \citenamefont {Chan}, \citenamefont {Aljunid}, \citenamefont {Shakhmuratov},\
  and\ \citenamefont {Wilkowski}}]{KCA19}%
  \BibitemOpen
  \bibfield  {author} {\bibinfo {author} {\bibfnamefont {C.~C.}\ \bibnamefont
  {Kwong}}, \bibinfo {author} {\bibfnamefont {E.~A.}\ \bibnamefont {Chan}},
  \bibinfo {author} {\bibfnamefont {S.~A.}\ \bibnamefont {Aljunid}}, \bibinfo
  {author} {\bibfnamefont {R.}~\bibnamefont {Shakhmuratov}},\ and\ \bibinfo
  {author} {\bibfnamefont {D.}~\bibnamefont {Wilkowski}},\ }\bibfield  {title}
  {\bibinfo {title} {Large optical depth frequency modulation spectroscopy},\
  }\href {https://doi.org/10.1364/OE.27.032323} {\bibfield  {journal} {\bibinfo
   {journal} {Opt. Express}\ }\textbf {\bibinfo {volume} {27}},\ \bibinfo
  {pages} {32323} (\bibinfo {year} {2019})}\BibitemShut {NoStop}%
\bibitem [{\citenamefont {Brandt}\ \emph {et~al.}(1997)\citenamefont {Brandt},
  \citenamefont {Nagel}, \citenamefont {Wynands},\ and\ \citenamefont
  {Meschede}}]{BNW97}%
  \BibitemOpen
  \bibfield  {author} {\bibinfo {author} {\bibfnamefont {S.}~\bibnamefont
  {Brandt}}, \bibinfo {author} {\bibfnamefont {A.}~\bibnamefont {Nagel}},
  \bibinfo {author} {\bibfnamefont {R.}~\bibnamefont {Wynands}},\ and\ \bibinfo
  {author} {\bibfnamefont {D.}~\bibnamefont {Meschede}},\ }\bibfield  {title}
  {\bibinfo {title} {Buffer-gas-induced linewidth reduction of coherent dark
  resonances to below $50 $hz},\ }\href
  {https://doi.org/10.1103/PhysRevA.56.R1063} {\bibfield  {journal} {\bibinfo
  {journal} {Phys. Rev. A}\ }\textbf {\bibinfo {volume} {56}},\ \bibinfo
  {pages} {R1063} (\bibinfo {year} {1997})}\BibitemShut {NoStop}%
\bibitem [{\citenamefont {Bouchiat}\ and\ \citenamefont
  {Brossel}(1966)}]{BB66}%
  \BibitemOpen
  \bibfield  {author} {\bibinfo {author} {\bibfnamefont {M.~A.}\ \bibnamefont
  {Bouchiat}}\ and\ \bibinfo {author} {\bibfnamefont {J.}~\bibnamefont
  {Brossel}},\ }\bibfield  {title} {\bibinfo {title} {Relaxation of optically
  pumped $\textrm{Rb}$ atoms on paraffin-coated walls},\ }\href
  {https://doi.org/10.1103/PhysRev.147.41} {\bibfield  {journal} {\bibinfo
  {journal} {Phys. Rev.}\ }\textbf {\bibinfo {volume} {147}},\ \bibinfo {pages}
  {41} (\bibinfo {year} {1966})}\BibitemShut {NoStop}%
\bibitem [{\citenamefont {Graf}\ \emph {et~al.}(2005)\citenamefont {Graf},
  \citenamefont {Kimball}, \citenamefont {Rochester}, \citenamefont {Kerner},
  \citenamefont {Wong}, \citenamefont {Budker}, \citenamefont {Alexandrov},
  \citenamefont {Balabas},\ and\ \citenamefont {Yashchuk}}]{GKR05}%
  \BibitemOpen
  \bibfield  {author} {\bibinfo {author} {\bibfnamefont {M.~T.}\ \bibnamefont
  {Graf}}, \bibinfo {author} {\bibfnamefont {D.~F.}\ \bibnamefont {Kimball}},
  \bibinfo {author} {\bibfnamefont {S.~M.}\ \bibnamefont {Rochester}}, \bibinfo
  {author} {\bibfnamefont {K.}~\bibnamefont {Kerner}}, \bibinfo {author}
  {\bibfnamefont {C.}~\bibnamefont {Wong}}, \bibinfo {author} {\bibfnamefont
  {D.}~\bibnamefont {Budker}}, \bibinfo {author} {\bibfnamefont {E.~B.}\
  \bibnamefont {Alexandrov}}, \bibinfo {author} {\bibfnamefont {M.~V.}\
  \bibnamefont {Balabas}},\ and\ \bibinfo {author} {\bibfnamefont {V.~V.}\
  \bibnamefont {Yashchuk}},\ }\bibfield  {title} {\bibinfo {title} {Relaxation
  of atomic polarization in paraffin-coated cesium vapor cells},\ }\href
  {https://doi.org/10.1103/PhysRevA.72.023401} {\bibfield  {journal} {\bibinfo
  {journal} {Phys. Rev. A}\ }\textbf {\bibinfo {volume} {72}},\ \bibinfo
  {pages} {023401} (\bibinfo {year} {2005})}\BibitemShut {NoStop}%
\bibitem [{\citenamefont {Klein}\ \emph {et~al.}(2011)\citenamefont {Klein},
  \citenamefont {Hohensee}, \citenamefont {Phillips},\ and\ \citenamefont
  {Walsworth}}]{KHP11}%
  \BibitemOpen
  \bibfield  {author} {\bibinfo {author} {\bibfnamefont {M.}~\bibnamefont
  {Klein}}, \bibinfo {author} {\bibfnamefont {M.}~\bibnamefont {Hohensee}},
  \bibinfo {author} {\bibfnamefont {D.~F.}\ \bibnamefont {Phillips}},\ and\
  \bibinfo {author} {\bibfnamefont {R.~L.}\ \bibnamefont {Walsworth}},\
  }\bibfield  {title} {\bibinfo {title} {Electromagnetically induced
  transparency in paraffin-coated vapor cells},\ }\href
  {https://doi.org/10.1103/PhysRevA.83.013826} {\bibfield  {journal} {\bibinfo
  {journal} {Phys. Rev. A}\ }\textbf {\bibinfo {volume} {83}},\ \bibinfo
  {pages} {013826} (\bibinfo {year} {2011})}\BibitemShut {NoStop}%
\bibitem [{\citenamefont {Fleischhauer}\ \emph {et~al.}(2000)\citenamefont
  {Fleischhauer}, \citenamefont {Matsko},\ and\ \citenamefont
  {Scully}}]{FMS00}%
  \BibitemOpen
  \bibfield  {author} {\bibinfo {author} {\bibfnamefont {M.}~\bibnamefont
  {Fleischhauer}}, \bibinfo {author} {\bibfnamefont {A.~B.}\ \bibnamefont
  {Matsko}},\ and\ \bibinfo {author} {\bibfnamefont {M.~O.}\ \bibnamefont
  {Scully}},\ }\bibfield  {title} {\bibinfo {title} {Quantum limit of optical
  magnetometry in the presence of ac stark shifts},\ }\href
  {https://doi.org/10.1103/PhysRevA.62.013808} {\bibfield  {journal} {\bibinfo
  {journal} {Phys. Rev. A}\ }\textbf {\bibinfo {volume} {62}},\ \bibinfo
  {pages} {013808} (\bibinfo {year} {2000})}\BibitemShut {NoStop}%
\bibitem [{\citenamefont {Budker}\ \emph {et~al.}(2002)\citenamefont {Budker},
  \citenamefont {Gawlik}, \citenamefont {Kimball}, \citenamefont {Rochester},
  \citenamefont {Yashchuk},\ and\ \citenamefont {Weis}}]{BGK02}%
  \BibitemOpen
  \bibfield  {author} {\bibinfo {author} {\bibfnamefont {D.}~\bibnamefont
  {Budker}}, \bibinfo {author} {\bibfnamefont {W.}~\bibnamefont {Gawlik}},
  \bibinfo {author} {\bibfnamefont {D.~F.}\ \bibnamefont {Kimball}}, \bibinfo
  {author} {\bibfnamefont {S.~M.}\ \bibnamefont {Rochester}}, \bibinfo {author}
  {\bibfnamefont {V.~V.}\ \bibnamefont {Yashchuk}},\ and\ \bibinfo {author}
  {\bibfnamefont {A.}~\bibnamefont {Weis}},\ }\bibfield  {title} {\bibinfo
  {title} {Resonant nonlinear magneto-optical effects in atoms},\ }\href
  {https://doi.org/10.1103/RevModPhys.74.1153} {\bibfield  {journal} {\bibinfo
  {journal} {Rev. Mod. Phys.}\ }\textbf {\bibinfo {volume} {74}},\ \bibinfo
  {pages} {1153} (\bibinfo {year} {2002})}\BibitemShut {NoStop}%
\end{thebibliography}%

\end{document}